\documentclass[prd,twocolumn,floats,floatfix,nofootinbib]{revtex4}
\usepackage[dvips]{graphicx}
\usepackage{graphics}
\usepackage{dcolumn}
\usepackage{amssymb}
\usepackage{bm}
\usepackage{amsmath}
\usepackage{color}

\def\spose#1{\hbox to 0pt{#1\hss}}

\def\lta{\mathrel{\spose{\lower 3pt\hbox{$\mathchar"218$}}
     \raise 2.0pt\hbox{$\mathchar"13C$}}}
\def\gta{\mathrel{\spose{\lower 3pt\hbox{$\mathchar"218$}}
     \raise 2.0pt\hbox{$\mathchar"13E$}}}
\newcommand{\be}{\begin{equation}}
\newcommand{\en}{\end{equation}}
\newcommand{\bea}{\begin{eqnarray}}
\newcommand{\ena}{\end{eqnarray}}

\begin{document}

\title{Ho\v{r}ava-Lifshitz Bouncing Bianchi IX Universes: A Dynamical System Analysis}

\author{Rodrigo Maier$^1$ and Ivano Dami\~ao Soares$^2$}

\affiliation{$^1$ Departamento de F\'isica Te\'orica, Instituto de F\'isica, Universidade do Estado do Rio de Janeiro,\\
Rua S\~ao Francisco Xavier 524, Maracan\~a, CEP 20550-900, Rio de Janeiro, Brasil,}

\affiliation{
$^2$Centro Brasileiro de Pesquisas F\'{\i}sicas -- CBPF, \\ Rua Dr. Xavier Sigaud, 150, Urca,
CEP 22290-180, Rio de Janeiro, Brazil}

\date{\today}

\begin{abstract}

We examine the Hamiltonian dynamics of bouncing Bianchi IX cosmologies with three scale factors
in Ho\v{r}ava-Lifshitz (HL) gravity. We assume a positive cosmological constant
plus noninteracting dust and radiation as the matter content of the
models. In this framework the modified field equations contain additional terms which turn the
dynamics nonsingular. The $6$-dim phase space presents (i) two critical points in a finite region of the
phase space, (ii) one asymptotic de Sitter attractor at infinity and
(iii) a $2$-dim invariant plane containing the critical points; together they organize the
dynamics of the phase space. We identified four distinct parameter domains $A$, $B$, $C$ and $D$
for which the pair of critical points engenders distinct features in the dynamics, connected
to the presence of centers of multiplicity two and saddles of multiplicity two.
In the domain $A$ the dynamics consists basically of periodic bouncing orbits, or oscillatory
orbits with a finite number of bounces before escaping to the de Sitter attractor.
The center with multiplicity two engenders in its neighborhood the topology of stable and
unstable cylinders $R \times S^3$ of orbits, where $R$ is a saddle direction and $S^3$ is
the center manifold of unstable periodic orbits. We show that the stable and
unstable cylinders coalesce realizing a smooth homoclinic connection to the center manifold,
a rare event of regular/non-chaotic dynamics in bouncing Bianchi IX cosmologies.
The presence of a saddle of multiplicity two in the domain $B$ engenders a high instability
in the dynamics so that the cylinders emerging from the center manifold
about $P_2$ towards the bounce have four distinct attractors: the center manifold itself,
the de Sitter attractor at infinity and two further momentum-dominated attractors
with infinite anisotropy.
In the domain $C$ we examine the features of invariant manifolds of orbits about a
saddle of multiplicity two $P_2$. The presence of the saddle of multiplicity two engenders
bifurcations of the invariant manifold as the energy $E_0$ of the system
increases relative to the energy $E_{cr_2}$ of $P_2$: (i) for $E_0 < E_{cr_2}$ the invariant
manifold has the topology $S^3$; (ii) for $E_0 = E_{cr_2}$ two points of $S^3$ pinch into the point
$P_2$, so that the invariant manifold contains infinitely many orbits homoclinic to $P_2$;
(iii) for $E_0 > E_{cr_2}$ the center manifold bifurcates into a $3$-torus; (iv) for $E_0$
sufficiently large the $3$-torus bifurcates into three $S^3$, an invariant manifold multiply connected.
Such structures were not yet observed in the literature.
The domain $D$ is not examined as most of its features are present already in the previous domains.


\end{abstract}

\maketitle

\section{Introduction}

Although General Relativity is the most successful theory that currently describes
gravitation, it presents some intrinsic crucial pathologies when one tries to construct a
cosmological model of a proper theory of gravitation. In cosmology, the $\Lambda {\rm CDM}$ model gives us important predictions concerning
the evolution of the Universe and about its current state \cite{mukhanov,weinberg,gorbunov, peter}. However, let us
assume that the initial conditions of our Universe were fixed when the early
Universe emerged from the semi-Planckian regime and started its classical
expansion. Evolving back such initial conditions using the Einstein field
equations, we see that our Universe is driven towards an initial singularity
where the classical regime is no longer valid \cite{wald}.
\par
Notwithstanding the cosmic censorship conjecture \cite{penrose}, there is no doubt
that General Relativity must be properly corrected or even replaced by a
completely new theory, let us say a quantum theory of gravity. This demand
is in order to solve the issue of the presence of the initial singularity predicted by
classical General Relativity in the beginning of the
Universe.
\par
One of the most important characteristics of our Universe supported by
observational data is its homogeneity and isotropy at large scales. However, when we consider
a homogeneous and isotropic model filled with baryonic matter, we find
several difficulties by taking into account the primordial state of our
Universe. Among such difficulties, we can mention the horizon and flatness
problems \cite{mukhanov,weinberg,gorbunov, peter}. Although the Inflationary
Paradigm\cite{abbott} allows one to solve problems like these, inflationary cosmology does not solve the problem of the initial
singularity.
\par
On the other hand,
since 1998 \cite{riess} observational data have been giving support to the
highly unexpected assumption that our Universe is currently in a state of accelerated expansion.
In order to explain this state of late-time acceleration, cosmologists have been considering the existence of some field
-- known as dark energy -- that violates the strong energy condition. Although it poses a problem to quantum
field theory on how to accommodate its observed value
with vacuum energy calculations\cite{ccp}, the cosmological constant seems to be the simplest and most
appealing candidate for dark energy. Therefore, nonsingular models which provide late-time acceleration should be strongly considered.
\par
During the last decades, bouncing models \cite{novello,peter1,peter2,ns} have been considered in
order to solve the problem of initial singularity predicted by General
Relativity. Such models (as in \cite{nelson,rm1,rm2}) might provide attractive alternatives
to the inflationary paradigm
once they can solve the horizon and flatness problems, and justify the
power spectrum of primordial cosmological perturbations
inferred by observations.
\par
In 2009, P. Ho\v{r}ava proposed a
modified gravity theory by considering a Lifshitz-type anisotropic
scaling between space and time at high energies \cite{Horava}.
In this context, it has been shown \cite{cal, brand} that higher
spatial curvature terms can lead to regular bounce solutions in
the early Universe. Since its proposal, several versions of Ho\v{r}ava-Lifshitz
gravity have emerged.
\par
In the case of a $4$-dimensional ($1+3$) spacetime, the basic assumption which
is required by all the versions of Ho\v{r}ava-Lifshitz theories is
that a preferred foliation of spacetime
is {\it a priori} imposed. Therefore it is
natural to work with the Arnowitt-Deser-Misner (ADM) decomposition
of spacetime
\begin{eqnarray}
\label{eq1}
ds^2=N^2dt^2-{^{(3)}g}_{ij}(N^i dt+dx^i)(N^j dt+dx^j),
\end{eqnarray}
where $N=N(t,x^i)$ is the lapse function, $N^i=N^i(t,x^i)$ is the shift and ${^{(3)}g}_{ij}={^{(3)}g}_{ij}(t,x^i)$ is the spatial geometry. In this case the
final action of the theory will not be invariant under diffeomorphisms as in General Relativity. Nevertheless,
an invariant foliation preserving diffeomorphisms can be assumed. This is achieved if the action is invariant
under the symmetry of time reparametrization together with time-dependent spatial diffeomorphisms. That is:
\begin{eqnarray}
\label{eq2}
t\rightarrow \bar{t}(t),~x^i\rightarrow\bar{x}^i(t,x^i).
\end{eqnarray}
It turns out that the only covariant object under spatial diffeomorphisms
that contains one time derivative of the spatial metric is the extrinsic curvature $K_{ij}$
\begin{eqnarray}
\label{eq3}
K_{ij}=\frac{1}{2N}\Big[\frac{\partial {^{(3)}g}_{ij}}{\partial t}-\nabla_i N_j-\nabla_j N_i\Big]
\end{eqnarray}
where $\nabla_i$ is the covariant derivative built with the spatial metric ${^{(3)}g}_{ij}$.
Thus, to construct the general theory which is of second order in time derivatives, one needs to consider
the quadratic terms $K_{ij} K^{ij}$ and $K^2$ -- where $K$ is the trace of $K_{ij}$ -- in the extrinsic curvature.
By taking these terms into account we obtain the following general action
\begin{eqnarray}
\label{eq4}
\nonumber
S_{\rm HL}\propto \int N \sqrt{{^{(3)}g}}\Big[(K_{ij} K^{ij} - \lambda K^2 - {^{(3)}R})\\
 - U_{\rm HL}({^{(3)}g}_{ij},N)\Big] d^3x dt
\end{eqnarray}
where ${^{(3)}g}$ is the
determinant of the spatial metric and $\lambda$ is a constant which corresponds to a dimensionless running coupling.
As in General Relativity the term $K_{ij}K^{ij}-K^2$ is invariant under four-dimensional diffeomorphisms,
we expect to recover the classical regime for $\lambda\rightarrow 1$. That is why it is a consensus that $\lambda$
must be a parameter sufficiently close to $1$.
In general, $U({^{(3)}g}_{ij},N)$ can depend on the spatial metric and the lapse function because of the symmetry of the theory.
It is obvious that there are several invariant terms that one could include in $U$.
Particular choices resulted in different versions of Ho\v{r}ava-Lifshitz gravity.
\par
Motivated by condensed matter systems, P. Ho\v{r}ava proposed a symmetry on $U$ that substantially reduces the
number of invariants\cite{Horava}. In this case, $U$ depends on a superpotential W given by the Chern-Simons term,
the curvature scalar and a term which mimics the cosmological constant.
It has been shown \cite{prob1} that this original assumption has to be broken if one intends to build a theory in
agreement to current
observations.
\par
The simplification $N=N(t)$ was also originally proposed by Ho\v{r}ava\cite{Horava}.
This condition defines a version of Ho\v{r}ava-Lifshitz gravity called Projectable. As $\partial N / \partial x^i \equiv 0$, the Projectable version also reduces the number of invariants that one can include in $U$.
The linearization of this version assuming a Minkowski background provides an extra scalar degree of freedom which
is classically unstable in the IR when $\lambda>1$ or $\lambda<1/3$, and is a ghost when $1/3<\lambda<1$ \cite{sotiriou2}. Although some physicists argue that
higher order derivatives can cut off these instabilities, it has been shown\cite{prob1,blas,bog,koyama} that a perturbative analysis is not consistent when $\lambda\rightarrow 1$ and the scalar mode gets strongly coupled. That is because the strongly coupled scale is unacceptably low. In this case, higher
order operators would modify the graviton dynamics at very low energies, being in conflict with current observations.
\par
Besides pure curvature invariants of ${^{(3)}g}_{ij}$, one may also include invariant contractions of $a_i\equiv\partial (\ln{N})/\partial x^i$ in $U$. This assumption defines
the so-called Non-Projectable version of Ho\v{r}ava-Lifshitz gravity. Connected to the lowest order invariant $a_i a^i$, there is a parameter $\sigma$ which defines a ``safe'' domain of the theory\cite{sotiriou2,sotiriou1}. In fact, in this case there is also an extra scalar degree of freedom when one linearizes the theory
in a Minkowski background. However, when $0<\sigma<2$ and $\lambda>1$ this mode is not a ghost nor classically unstable (as long as detailed balance is not imposed).
Although the Non-Projectable version also exhibits a strong coupling\cite{prob1,sotiriou1,pap}, it has been argued that its scale is too high to be phenomenologically accessible from gravitational experiments\cite{sotiriou2}.
\par
In this paper we adhere to a particular version of Non-Projectable Ho\v{r}ava-Lifshitz gravity,
in which the potential $U_{\rm HL}$ is given by
\begin{eqnarray}
\label{eq5}
\nonumber
&&U_{\rm HL}= \sigma a_i a^{i}+\alpha_{21}~ ^{(3)}R^2+ \alpha_{22}~ ^{(3)}R^{i}_{j}~ ^{(3)}R^{j}_{i}+ \alpha_{31}~ ^{(3)}R^{3}\\
&+&\alpha_{32}~ ^{(3)}R~ ^{(3)}R^{i}_{j}~ ^{(3)}R^{j}_{i} + \alpha_{33}~ ^{(3)}R^{i}_{j}~ ^{(3)}R^{j}_{k}~^{(3)}R^{k}_{i},
\end{eqnarray}
where ${^{(3)}R}_{ij}$ is the spatial Ricci tensor and $\alpha_{ij}$ are coupling constants. To complete the above
Ho\v{r}ava-Lifshitz action we add the remaining action
\begin{eqnarray}
S\propto \int N \sqrt{{^{(3)}g}}[-2 \Lambda -2 {\cal L}_m ] d^3x dt,
\label{lagr2}
\end{eqnarray}
with a cosmological constant $\Lambda$ and where ${\cal{L}}_m$ is the Lagrangean density
of the matter content of the model, which we take as dust and radiation.
\par
In the next section we analyze the structure of the phase space of a nonsingular
Bianchi IX cosmological model with three scale factors
-- sourced with dust, radiation and a cosmological constant --
which arises from Non-Projectable Ho\v{r}ava-Lifshitz gravity.
\par
A similar model was previously considered by Misonoh, Maeda and Kobayashi
\cite{maeda} and analyzed numerically. However their work did not contemplate
the full Hamiltonian formulation of the phase space of the system and its basic
and fundamental structures that organize the dynamics in the whole phase space.
The connection of the authors' results with ours are discussed in the paper.
{\color{black}For future reference we mention here that the parameters of the
potential $\mathcal{V}_{HL}$ used in \cite{maeda}, eq. (2.6), are related to the
corresponding parameters of our paper according to
\begin{eqnarray}
\nonumber
&&g_{2}=\alpha_{31}, ~~g_{3}=\alpha_{22},\\
\nonumber
&&g_{5}=\alpha_{31}, ~~g_{6}=\alpha_{32},~~g_{7}=\alpha_{33}.
\end{eqnarray}
The choice $g_1=-1$ in \cite{maeda} is equivalent to include
$^{(3)}R$ in the expression $(K_{ij}K^{ij}- \lambda K^2 - ^{(3)}R)$ of eq. (\ref{eq4}) of our paper;
for $\lambda=1$ this expression constitutes the gravitational action of GR in the ADM formalism.
The parameter $g_8$ multiplies an expression that is zero in the case of the spatially homogeneous Bianchi IX metric;
therefore this term was not included in $U_{HL}$, eq. (\ref{eq5}) above. Concerning $g_4$ and $g_9$ we
did not consider $HL$ potential terms containing covariant spatial derivatives of the 3-dim Ricci
tensor $^{(3)}R_{ij}$. The non-canonical variables $(a,\beta_{+},\beta_{-})$ of \cite{maeda} are
related to the canonical variables $(x,y,z)$, defined in Section \ref{sectiona}, by
\begin{eqnarray}
\nonumber
a=2x,~~ \beta_{+}=(\ln z) /6,~~\beta_{-}=\sqrt{3}~(\ln y)/6.
\end{eqnarray}
}
\section{The Model}

\par
The fundamental symmetry assumed in Ho\v{r}ava-Lifshitz gravity provides enough gauge freedom to choose
\begin{eqnarray}
\label{eq8a}
N=N(t),~~ N_i=0.
\end{eqnarray}
Let us then consider a general Bianchi IX spatially homogeneous geometry with  three scale factors
in comoving coordinates,
\begin{eqnarray}
\label{eq2}
ds^2=N^2~dt^2+h_{ij} \omega^{i}\omega^{j}
\label{ds}
\end{eqnarray}
where $t$ is the cosmological time and
\begin{eqnarray}
\label{eq8a1}
\nonumber
h_{ij}&=&{\rm diag}(-M^2,-Q^2,-R^2),\\
h^{ij}&=&{\rm diag}(-\frac{1}{M^2},-\frac{1}{Q^2},-\frac{1}{R^2}).
\end{eqnarray}
$(M(t),Q(t),R(t))$ are the scale factors of the model in the
Bianchi IX 1-form basis $\omega^{i}$ ($i=1,2,3$) which satisfy
\begin{eqnarray}
\label{eq4ñ}
d\omega^{i}= \frac{1}{2}\epsilon^{ijk} \omega^{j} \wedge\omega^{k},
\end{eqnarray}
where $d$ denotes the exterior derivative. In the basis $\omega^{i}$ we have
\begin{eqnarray}
\label{eq8a1}
\nonumber
K_{ij}=-\frac{1}{N} {\dot{h}_{ij}}=\frac{1}{N}(-M {\dot{M}},-Q{\dot{Q}},-R{\dot{R}}),
\end{eqnarray}
and
\begin{eqnarray}
\label{eq8a1}
K^{ij}=-\frac{1}{N} {\dot{h}_{ij}}=\frac{1}{N}(-\frac{\dot{M}}{M^3},-\frac{\dot{Q}}{Q^3},-\frac{\dot{R}}{R^3}),
\label{eq8a2}
\end{eqnarray}
For future reference the nonvanishing spatial components of ${^{(3)}R}^i_{~j}$ are given by
\begin{eqnarray}
\label{eq8c0}
\nonumber
{^{(3)}R}^1_{~1}&=&-\frac{1}{M^2}+\frac{1}{2}\Big[-\frac{M^2}{Q^2R^2}+\frac{Q^2}{M^2R^2}+\frac{R^2}{M^2Q^2}\Big]
\\
\nonumber
{^{(3)}R}^2_{~2}&=&-\frac{1}{Q^2}+\frac{1}{2}\Big[\frac{M^2}{Q^2R^2}-\frac{Q^2}{M^2R^2}+\frac{R^2}{M^2Q^2}\Big]
\\
\nonumber
{^{(3)}R}^3_{~3}&=&-\frac{1}{R^2}+\frac{1}{2}\Big[\frac{M^2}{Q^2R^2}+\frac{Q^2}{M^2R^2}-\frac{R^2}{M^2Q^2}\Big]
\end{eqnarray}
so that
\begin{eqnarray}
\label{eq8f}
\nonumber
{^{(3)}}R=\frac{1}{2M^2Q^2R^2}[M^4+Q^4+R^4-(R^2-Q^2)^2\\
-(R^2-M^2)^2-(M^2-Q^2)^2],
\end{eqnarray}
\begin{eqnarray}
\label{eq8f1}
\nonumber
{^{(3)}}R^i_{~j}{^{(3)}}R^j_{~i}=\frac{1}{4(MQR)^4}[3M^8-4M^6(Q^2+R^2)\\
\nonumber
-4M^2(Q^2-R^2)^2(Q^2+R^2)+2M^4(Q^2+R^2)^2\\
+(Q^2-R^2)^2(3Q^4+2Q^2R^2+3R^4)],
\end{eqnarray}
and
\begin{eqnarray}
\label{eq8f2}
\nonumber
{^{(3)}}R^i_{~j}{^{(3)}}R^j_{~k}{^{(3)}}R^k_{~i}=\frac{1}{8(MQR)^6}\{[(M^2-Q^2)^2-R^4]^3
\\
+[(M^2-R^2)^2-Q^4]^3+[(Q^2-R^2)^2-M^4]^3\},~~~~~
\end{eqnarray}
which are the key terms to evaluate the potential $U_{\rm HL}$.
Therefore, Lagrangian of the total action resulting from (\ref{eq4})-(\ref{lagr2}) is given,
up to a constant volume integral, by
\begin{eqnarray}
\label{eq8c}
\nonumber
{\cal L} \propto K - V
\end{eqnarray}
where the kinetic part $K$ is given by
\begin{eqnarray}
\nonumber
K=\frac{MQR}{N}\Big[{(1-\lambda)}\Big(\frac{\dot{M}^2}{M^2}+\frac{\dot{Q}^2}{Q^2}+\frac{\dot{R}^2}{R^2}\Big)\\
- {2}\lambda\Big(\frac{\dot{M}\dot{Q}}{MQ}+\frac{\dot{Q}\dot{R}}{QR}+\frac{\dot{M}\dot{R}}{MR}\Big)\Big],
\label{eqK}
\end{eqnarray}
and the potential part $V$ is
\begin{eqnarray}
\label{eqV}
\nonumber
V= - N (MQR)\Big[ {^{(3)}R} +2 \Lambda + U_{\rm HL}\\
+2  \Big(E_0 +\frac{E_r}{(MQR)^{1/3}} \Big) \Big],
\end{eqnarray}
where $E_0$ and $E_r$ are constants, corresponding to the separately conserved
total energy of dust and radiation, respectively. $U_{\rm HL}$ was fixed in (\ref{eq5}).
\par By defining then the canonical momenta as
\begin{eqnarray}
\label{eq8d}
p_M=\frac{\partial {\cal L}}{\partial \dot{M}},~p_Q=\frac{\partial {\cal L}}{\partial \dot{Q}},~
p_R=\frac{\partial {\cal L}}{\partial \dot{R}},
\end{eqnarray}
the total action can be reexpressed as
\begin{eqnarray}
\label{eqH}
S \propto \int \Big(\sum_{i} {\dot q}_i  p_i -N {\cal H} \Big) dt
\end{eqnarray}
so that $\delta S/\delta N =0$ results in the first integral of motion,
the conserved Hamiltonian constraint
\begin{eqnarray}
\label{eq8e}
\nonumber
{\cal H}=\frac{1}{4(3\lambda-1)}\Big[(2\lambda-1)\Big(\frac{M p^2_M}{QR}+\frac{Q p^2_Q}{MR}+\frac{R p^2_R}{MQ}
\Big)\\
\nonumber
- 2\lambda\Big(\frac{p_M p_Q}{R}+\frac{p_M p_R}{Q}+\frac{p_Q p_R}{M}\Big)\Big]+2\Lambda MQR\\
+2E_0+\frac{ 2E_r}{(MQR)^{\frac{1}{3}}}+MQR[{^{(3)}}R+U_{\rm HL}]=0.
\end{eqnarray}
From the point of view of dynamical systems we may consider $E_0$ in (\ref{eq8e}) as the total
conserved energy of the Hamiltonian dynamics so that we will refer to it as the total energy of the system.
We also assume a positive cosmological constant $\Lambda>0$.
\par
From (\ref{eq8e}) we derive the equations of motion
\begin{widetext}
\begin{eqnarray}
\label{eq9}
\nonumber
\dot{M}&=&\frac{(1-2\lambda)Mp_M+\lambda(Qp_Q+Rp_R)}{2QR(1-3\lambda)}\\
\nonumber
\dot{Q}&=&\frac{(1-2\lambda)Qp_Q+\lambda(Mp_M+Rp_R)}{2MR(1-3\lambda)}\\
\nonumber
\dot{R}&=&\frac{(1-2\lambda)Rp_R+\lambda(Mp_M+Qp_Q)}{2QM(1-3\lambda)}\\
{\dot{p}_M}&=&\frac{(1-2\lambda)[M^2p^2_M+Q^2p^2_Q+R^2p^2_R]-2\lambda QRp_Qp_R}{4(3\lambda-1)M^2QR}
-\Lambda QR +\frac {QR E_r}{3(MQR)^\frac{4}{3}}\\
\nonumber
&+&QR [{^{(3)}}R+U]
+MQR\frac{\partial}{\partial M}[{^{(3)}}R+U_{\rm HL}]\\
\nonumber
{\dot{p}_Q}&=&\frac{(1-2\lambda)[M^2p^2_M+Q^2p^2_Q+R^2p^2_R]-2\lambda MRp_Mp_R}{4(3\lambda-1)MQ^2R}
-\Lambda MR +\frac {MR E_r}{3(MQR)^\frac{4}{3}}\\
\nonumber
&+&MR [{^{(3)}}R+U]
+MQR\frac{\partial}{\partial Q}[{^{(3)}}R+U_{\rm HL}]~~~~\\
\nonumber
{\dot{p}_R}&=&\frac{(1-2\lambda)[M^2p^2_M+Q^2p^2_Q+R^2p^2_R]-2\lambda MQp_Mp_Q}{4(3\lambda-1)MQR^2}
\nonumber
-\Lambda MQ +\frac {MQ E_r}{3(MQR)^\frac{4}{3}}\\
\nonumber
&+&MQ [{^{(3)}}R+U]
+MQR\frac{\partial}{\partial R}[{^{(3)}}R+U_{\rm HL}].~~~~
\end{eqnarray}
\end{widetext}
The above equations were derived for the most general case in which $\lambda$ is
an additional free parameter of the model.
From a dynamical system point of view, this would be interesting in order to study
the role of $\lambda$ in the phase space dynamics.
However, in order to recover General Relativity in the IR, not only $\sigma \rightarrow 0$,
but also $\lambda \rightarrow 1$\cite{sotiriou2}.
In fact, in the framework of Ho\v{r}ava-Lifshitz, $\lambda$ must be sufficiently close to $1$
in order to guarantee that no serious Lorentz invariance violation occurs. Therefore,
in order to simplify our analysis, in the remaining of the paper we will be restricted to the case $\lambda=1$.
\section{The Skeleton of the Phase Space\label{section3}}
In order to have an overall view of the phase space of the system, in the present section we will
examine the basic structures that organize the dynamics of the phase space. The first of these is
the invariant plane defined by
\begin{eqnarray}
\label{eq08091}
p_M=p_Q=p_R,~~~~~~M=Q=R,
\end{eqnarray}
so that the Hamiltonian (\ref{eq8e}) for the dynamics in the invariant plane reduces to
\begin{eqnarray}
\label{eq08092}
{\cal H}_{I}=\frac{3}{8} \frac{p_M^2}{M}+V(M)-2 E_0=0.~~
\end{eqnarray}
where
\begin{eqnarray}
\label{eq08093}
V(M)&=&\frac{3}{2}M-2 \Lambda M^3-\frac{A_2}{M}+\frac{A_3}{M^3},\\
\label{eqC1}
A_2&=&3\alpha_{21}+ \alpha_{22}+\frac{8 E_r}{3},\\
\label{eqC2}
A_3&=&9\alpha_{31}+3 \alpha_{32}+ \alpha_{33}.
\end{eqnarray}
From the expression of $V(M)$ we see that the bounce condition implies
$A_3>0$, so that we will restrict ourselves to this case in the paper,
corresponding to a well-behaved dynamics. Furthermore in order to have
a deSitter attractor at infinity, corresponding to a possible exponentially
expanding phase for orbits of the system, we will fix $\Lambda>0$.
\par The critical points of the phase space are
defined as equilibrium points of the dynamics (\ref{eq9}), and given by
\begin{eqnarray}
\label{eq10}
p_M=p_Q=p_R=0,~~~~~~M=Q=R=M_0,
\end{eqnarray}
where $M_0$ is a positive constant satisfying
\begin{eqnarray}
\label{eq11}
M_0^6-\frac{M_0^4}{4\Lambda}-\frac{A_2 M_0^2}{8\Lambda}+\frac{3A_3}{16\Lambda}=0,
\end{eqnarray}
so that the right-hand-side of (\ref{eq9}) vanishes. Obviously the critical points belong
to the invariant plane. From (\ref{eq08092}) we obtain that
the energy of a critical point $M_0$ is given by
\begin{eqnarray}
\label{eq11i}
E_{cr}= \frac{3}{4}M_0- \Lambda M_0^3+\frac{3}{16} \frac{A_3}{M_0^3}-\frac{3}{8} \frac{A_2}{M_0}.
\end{eqnarray}
By fixing $A_3>0$ and $\Lambda>0$ as postulated above, a careful analysis of (\ref{eq11})
shows that we have at most two critical points, or one critical point or no critical point
depending on the values of $\Lambda$, $A_2$ and $A_3$. Each critical point corresponds to a real
positive root of (\ref{eq11}) with $E_{cr}>0$ in (\ref{eq11i}).
\begin{figure}
\includegraphics[width=8cm,height=6cm]{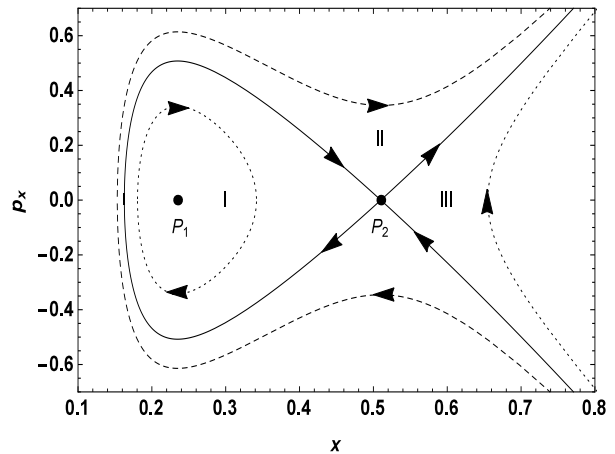}
\caption{The invariant plane. Here we fixed the parameters $\Lambda=1$, $A_2=0.05$ and $A_3=0.005$.
Dashed, solid and dotted lines correspond to $E_0=0.250$, $E_0=0.2201517192605279$ and $E_0=0.185$, respectively.
The second value of the energy is the energy of the critical point $P_2$, and the solid line constitutes
a homoclinic connection of $P_2$ to itself.
The critical points are given by $P_1=(0.2348551826828089,0)$ and $P_2=(0.51007113736321,0)$.
The graph was made in the canonical variables $(x,p_x)$ introduced in Section \ref{sectiona}.}
\label{detM}
\end{figure}
\par Figure \ref{detM} illustrates the invariant plane and the critical points in the
finite region of the phase space for the parameters $\Lambda=1$, $A_2=0.05$ and $A_3=0.005$.
The graph is made in the canonical variables $(x,p_x)$ of the invariant plane to be introduced in Section
\ref{sectiona}. In these coordinates the critical points are given by $P_1=(0.2348551826828089,0)$
and $P_2=(0.51007113736321,0)$. Dashed and dotted orbits shown in the invariant plane correspond to the
energies $E_0=0.250$, and $E_0=0.185$, respectively. The energy of the critical point $P_2$ is $E_0=0.2201517192605279$
and corresponds also to the separatrix (solid line) which is homoclinic connection of
$P_2$ to itself. The separatrix divides the invariant plane into three disconnected regions:
region $(I)$, of bounded periodic orbits corresponding to eternal oscillating universe, and regions $(II)$
and $(III)$ of one-bounce universes emerging from the deSitter repeller and tending
to a deSitter attractor at infinity. The scale factor approaches the deSitter
asymptotic configurations as $x \sim \exp(t\sqrt{\Lambda/3})$ and $p_x \sim \exp(t\sqrt{4\Lambda/3})$ for
times going to $\pm \infty$. As will be shown along the paper some parametric configurations
may also present velocity dominated attractors at infinity.
\par Finally we should mention that the phase space of the dynamical system (\ref{eq9})
presents two invariant submanifolds defined by
\begin{eqnarray}
\label{eqInvSub1}
M=Q,~~~p_M=p_Q,
\end{eqnarray}
and
\begin{eqnarray}
\label{eqInvSub2}
Q=R,~~~p_Q=p_R.
\end{eqnarray}

The denomination invariant submanifolds derives from the fact that each of them is mapped
into itself by the general Hamiltonian flow (\ref{eq9}), in other words, is invariant under the flow.
In particular the invariant plane (\ref{eq08091}) corresponds to the intersection of these two submanifolds and
satisfies obviously this property.
\par The nature of the critical points is characterized by linearizing the dynamical
equations (\ref{eq9}) about the critical point. Defining
\begin{eqnarray}
\label{eq11ii}
&&X=(M - M_0),~~W=(p_M -0),\\
&&Y=(Q-M_0),~~K=(p_Q-0),\\
&&Z=(R-M_0),~~L=(p_R-0),
\end{eqnarray}
small, we obtain from (\ref{eq9})
\begin{eqnarray}
\label{eq99}
\left(
\begin{array}{c}
\dot{X}  \\
\dot{Y}   \\
\dot{Z}   \\
\dot{W}   \\
\dot{K}   \\
\dot{L}
\end{array}
\right)=\left(
\begin{array}{cccccc}
0 & 0 & 0 & \alpha & -\alpha & -\alpha  \\
0 & 0 & 0 & -\alpha & \alpha & -\alpha  \\
0 & 0 & 0 & -\alpha & -\alpha & \alpha  \\
\delta & \gamma & \gamma & 0 & 0 & 0 \\
\gamma & \delta & \gamma & 0 & 0 & 0 \\
\gamma & \gamma & \delta & 0 & 0 & 0
\end{array}
\right)
\left(
\begin{array}{c}
X  \\
Y   \\
Z   \\
W   \\
K   \\
L
,
\end{array}
\right)
\end{eqnarray}
where
\begin{eqnarray}
\label{eqLine2}
\alpha&=&\frac{1}{4M_{0}},\\
\nonumber
\delta&=&\frac{1}{M_{0}^3}(-\frac{8 E_r}{9}-3\alpha_{22}+7\alpha_{21})-\frac{3}{M_0}\\
&&+\frac{1}{4M_0^5}(27 \alpha_{33}-45\alpha_{31}+17 \alpha_{32}),\\
\nonumber
\gamma&=&\frac{1}{4 M_{0}^3}(-\frac{8 E_r}{9}+5 \alpha_{22}-17 \alpha_{21})+\frac{3}{2M_0}\\
&&+\frac{1}{8 M_0^5}(-21 \alpha_{33}+99 \alpha_{31}+\alpha_{32}) -2\Lambda M_{0}.
\end{eqnarray}
The nature of a critical points $M_0$ is determined by the characteristic polynomial
associated with the linearization matrix in (\ref{eq99}). We obtain
\begin{eqnarray}
\label{poly}
P(L)= (L-L_1)(L+L_1)(L-L_2)^2 (L+L_2)^2~,
\end{eqnarray}
with roots
\begin{eqnarray}
L_1=\pm \sqrt{-\alpha(2\gamma+\delta)},~~~~L_2=\pm \sqrt{2\alpha(\delta-\gamma)},
\end{eqnarray}
where the second pair has multiplicity two.
\par We see that the characterization of the critical points $M_0$ and of
the structure of the phase in its neighborhood of the critical points is highly complex, depending
on the domains of the parameters appearing in the Hamiltonian (\ref{eq8e}).
\par
With view to a numerical illustration we give here $L_1$ and $L_2$ in terms of the parameters,
\begin{eqnarray}
\label{L1}
&&L_1=\pm\frac{1}{2\sqrt{2}}\sqrt{8\Lambda+\frac{A_2}{M_0^4}-\frac{3A_3}{M_0^6}},\\
\label{L2}
\nonumber
&&L_2=\pm\frac{\sqrt{2}}{4M_0^3\sqrt{3}}\Big(\frac{225}{2}A_3-144(9\alpha_{31}+2\alpha_{32})-M_0^2\times\\
&&(51 A_2+54M_0^2-128E_r-288\alpha_{21}-24\Lambda M_0^4)\Big)^{1/2},
\end{eqnarray}
with $L_2$ having multiplicity $2$. As we are restricting ourselves to the case of two critical points,
namely $\Lambda>0$ and $A_3>0$, four main configurations are present. Let $P_1$ and $P_2$ denote the
two critical points in the invariant plane. The following possible configurations
are then:\\
$(A)$ $P_1$ is a center-center-center  and $P_2$ is a saddle-center-center;\\
$(B)$ $P_1$ is a center-saddle-saddle  and $P_2$ is a saddle-center-center;\\
$(C)$ $P_1$ is a center-saddle-saddle and $P_2$ is a saddle-saddle-saddle;\\
$(D)$ $P_1$ is a center-center-center and $P_2$ is a saddle-saddle-saddle.\\
In the above we must remark that the
denomination ``center-center-center'' actually denotes the topology of a center
times a center with multiplicity $2$, and ``saddle-saddle-saddle'' denotes the topology of a saddle
times a saddle with multiplicity $2$, and so on.
\par For an illustration of the parameter domains corresponding to such configurations
let us fix $A_2=0.05$, $A_3=0.005$ and $\Lambda=1$. Furthermore, we will also fix $E_r=0.1$.
We obtain for the four configurations:

\vspace{0.3cm}

(A) $P_1$ is a center-center-center and $P_2$ is a saddle-

center-center:
$\alpha_{21} < -0.0609122 + 81.5854 \alpha_{31}$

$+ 18.1301 \alpha_{32}$ and $\alpha_{21} < 0.0000442278 + 17.2962 \alpha_{31}$

$+ 3.8436 \alpha_{32}$,

\vspace{0.1cm}

(B) $P_1$ is a center-saddle-saddle and $P_2$ is a saddle-

center-center: $\alpha_{21} > -0.0609122 + 81.5854 \alpha_{31}$

$+ 18.1301 \alpha_{32}$ and $\alpha_{21} < 0.0000442278 + 17.2962 \alpha_{31}$

$+ 3.8436 \alpha_{32}$,

\vspace{0.1cm}

(C) $P_1$ is a center-saddle-saddle and $P_2$ is a saddle-

saddle-saddle: $\alpha_{21} > -0.0609122 + 81.5854 \alpha_{31}$

$+ 18.1301 \alpha_{32}$ and $\alpha_{21} > 0.0000442278 + 17.2962 \alpha_{31}$

$+ 3.8436 \alpha_{32}$,

\vspace{0.1cm}

(D) $P_1$ is a center-center-center and $P_2$ is a saddle-

saddle-saddle: $\alpha_{21} < -0.0609122 + 81.5854 \alpha_{31}$

$+ 18.1301 \alpha_{32}$ and $\alpha_{21} > 0.0000442278 + 17.2962 \alpha_{31}$

\begin{figure}
\includegraphics[width=8cm,height=6cm]{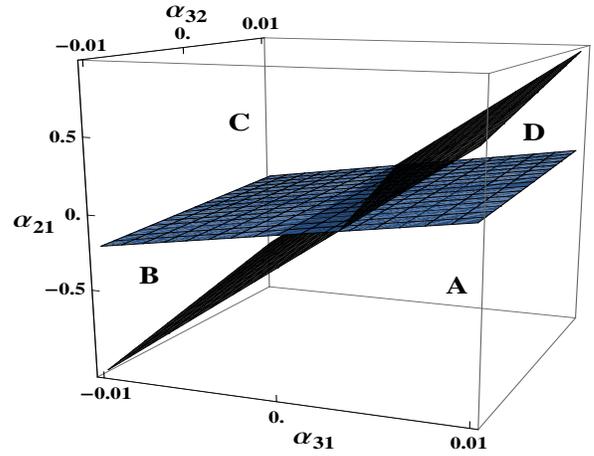}
\caption{The parameter space $(\alpha_{31},\alpha_{32},\alpha_{21})$ and the four
3-dim domains corresponding to the configurations $(A)$, $(B)$, $(C)$ and $(D)$
of the critical points.
}
\label{parameter}
\end{figure}

$+ 3.8436 \alpha_{32}$,

\vspace{0.3cm}

\noindent
which are illustrated in Fig. \ref{parameter}.
We must observe that the domains $(A)$, $(B)$, $(C)$ and $(D)$ do not
overlap by definition, and consequently the shaded surfaces shown in Fig. \ref{parameter}
do not belong to any of the four domains.
\par As we will see the high instability of the dynamics in
the cases $(B)-(D)$ is connected with the presence of saddles with multiplicity
two, discussed in the following sections.

\section{The dynamics about the critical points: the case of a center-center-center and a saddle-center-center\label{sectiona}}

We now describe the topology and the invariant manifolds of the dynamics in the linear
neighborhood of the critical points. We will then apply this analysis to the parameter domain
$(A)$, corresponding the case of a center-center-center $P_1$ and a
saddle-center-center $P_2$.
To start let us introduce the canonical transformation with the generating function
\begin{eqnarray}
\label{eqGen}
G=(M Q R)^{1/3} p_{x}+\frac{M}{Q} p_{y}+\frac{MQ}{R^2} p_{z},
\end{eqnarray}
where $p_{x}$, $p_{y}$ and $p_{z}$ are the new momenta, resulting in
\begin{eqnarray}
\label{eqGen1}
x=(M Q R)^{1/3},~~~y=\frac{M}{Q},~~~z=\frac{MQ}{R^2},
\end{eqnarray}
and
\begin{eqnarray}
\label{eqGen2}
\nonumber
p_M&=&\frac{1}{3}\frac{QR}{(MQR)^{2/3}}p_x+\frac{1}{Q}p_y+\frac{Q}{R^2}p_z,\\
p_Q&=&\frac{1}{3}\frac{MR}{(MQR)^{2/3}}p_x-\frac{M}{Q^2}p_y+\frac{M}{R^2}p_z,\\
\nonumber
p_R&=&\frac{1}{3}\frac{MQ}{(MQR)^{2/3}}p_x-\frac{2MQ}{R^3}p_z.
\end{eqnarray}
Here, the variable $x$ is obviously the average scale factor of the model.
In these new canonical variables the equations of the invariant plane reduce to
\begin{eqnarray}
\label{eqInv2}
y=1,~~~z=1,~~~p_y=0=p_z.
\end{eqnarray}
The new variables $(x,p_x)$ are then seen to be defined on the invariant plane. These variables
were used in constructing Fig. \ref{detM} displaying the invariant plane in the
parameter domain $(A)$.

\par In the new variables $(x,p_x,y,p_y,z,p_z)$ the full Hamiltonian (\ref{eq8e}) assumes the form
\begin{eqnarray}
\label{eq10091}
\nonumber
{\cal H}&=&-\frac{p_x^2}{24 x} + \frac{p_y^2 y^2}{2 x^3} + \frac{3 p_z^2 z^2}{2 x^3}+\frac{x}{2 z^{\frac{4}{3}}}
- \frac{x}{y z^\frac{1}{3}} - \frac{x y}{z^\frac{1}{3}} \\
\nonumber
&-& x z^\frac{2}{3} +
\frac{x z^\frac{2}{3}}{2 y^2} + \frac{1}{2} x y^2 z^\frac{2}{3} + 2 x^3 \Lambda +2 E_0 + \frac{2 E_r}{x}\\
&+&{x^3}U_{HL}(x,y,z)=0
\end{eqnarray}
\par These new canonical variables are very useful since they separate the degrees of freedom
of the system about the critical points into the expansion/contraction mode $(x,p_x)$,
connected to the invariant plane, and the modes $(y,p_y)$ and $(z,p_z)$ that -- in the case
of a center-center-center or a saddle-center-center -- are pure rotational modes about the critical point.
These variables allow us to describe the topology of the general dynamics in a linear
neighborhood of the critical points as well as to examine the nonlinear extension of invariant manifolds
about the critical points as we proceed to show.
\par To see this we expand the Hamiltonian (\ref{eq10091}) in a linear neighborhood of
the critical point $(x=M_0,p_x=0,y=1,p_y=0,z=1,p_y=0)$, resulting in the quadratic form
\begin{eqnarray}
\label{eqHH}
\nonumber
{\cal H}_L&=&2 (E_0-E_{cr})-\Big[\frac{p_x^2}{24 M_0} -q_x (x-M_0)^2 \Big]\\
\nonumber
&&+\Big[\frac{1}{2}\frac{p_y^2}{ M_0^3} + 3 q (y-1)^2\Big]\\
&&+\Big[\frac{3}{2} \frac{p_z^2}{M_0^3}+ q (z-1)^2\Big]=0,
\end{eqnarray}
where
\begin{eqnarray}
\label{eqHC}
\nonumber
q_x&=&6\Lambda M_0+\frac{1}{4M_0^3}({3\alpha_{22}}+{9 \alpha_{21}}+8 E_r)\\
&&-\frac{1}{4M_0^5}({9\alpha_{33}}+{27 \alpha_{32}}+{81 \alpha_{31}}),\\
\label{eqHC1}
\nonumber
q&=&\frac{1}{4M_0^3}({9\alpha_{31}}-{ \alpha_{32}}-3 \alpha_{33})\\
&&+\frac{1}{3M_0}({\alpha_{22}}-{3 \alpha_{21}})+\frac{M_0}{3}.
\end{eqnarray}
In deriving (\ref{eqHH}) the equations defining the critical points (\ref{eq11}) and their respective energy
(\ref{eq11i}) were used.
At this point it is worth mentioning that, in terms of the parameters $q_x$ and $q$, the four
parametric domain configurations given in the previous section can be simply characterized as
\begin{eqnarray}
\label{eqPAR}
\nonumber
(A) ~~~~P_1:~(q_x<0,~q>0),~~P_2:~(q_x>0,~q>0),\\
\nonumber
(B) ~~~~P_1:~(q_x<0,~q<0),~~P_2:~(q_x>0,~q>0),\\
\nonumber
(C) ~~~~P_1:~(q_x<0,~q<0),~~P_2:~(q_x>0,~q<0),\\
\nonumber
(D) ~~~~P_1:~(q_x<0,~q>0),~~P_2:~(q_x>0,~q<0).
\end{eqnarray}
\par
The quadratic Hamiltonian (\ref{eqHH}) is obviously separable and can be reexpressed as
\begin{eqnarray}
\label{eqHC2}
{\cal H}_L=2 (E_{cr}-E_0)+E_x - E_1 -E_2=0
\end{eqnarray}
where
\begin{eqnarray}
\label{eqHC3}
E_x&=&\frac{p_x^2}{24 M_0} -q_x (x-M_0)^2,\\
E_1&=&\frac{1}{2}\frac{p_y^2}{ M_0^3} + 3 q (y-1)^2,\\
E_2&=&\frac{3}{2} \frac{p_z^2}{M_0^3}+ q (z-1)^2,
\end{eqnarray}
are constants of motion of the linearized motion about the critical point $M_0$
in the sense that they have zero Poisson brackets with the Hamiltonian ${\cal H}_L$.
Two additional constants of the linearized motion are also present,
\begin{eqnarray}
\label{eqHC4}
C_1&=& \Big( \frac{1}{2 M_0^3} p_y p_z + q~ Y Z \Big),\\
\label{eqHC44}
C_2&=& \Big(Y p_z -\frac{1}{3} Z p_y \Big),
\end{eqnarray}
where in the above $Y \equiv (y-1)$ and $Z \equiv (z-1)$. They are  not
all independent but related by
\begin{eqnarray}
\label{eqHC5}
4 E_1 E_2= 12 C_1^2 + 6 C_2^2 .
\end{eqnarray}
We introduce a third constant of motion defined by
\begin{eqnarray}
\label{eqHC6-0}
C_3&=& \Big(E_1-E_2 \Big),
\end{eqnarray}
that together with $C_1$ and $C_2$ satisfy the algebra
\begin{eqnarray}
\label{eqHC6}
\nonumber
&&[C_1,C_2]=-\frac{1}{3}~C_3,~~~[C_2,C_3]=-4~ C_1,\\
&&[C_3,C_1]=-\frac{6 q}{M_0^3}~C_2.
\end{eqnarray}
\par
We are now ready to describe the topology of the 6-dim phase space in the linear neighborhood of the
critical points. In the remaining of this section we will restrict ourselves to
the parameter domain $(A)$ for which both critical points $P_1$ and $P_2$ have $q>0$, cf. (\ref{eqHC1}).
\par
Let us consider the case $E_x=0$ corresponding to $(x=M_0,p_x=0)$. The motion about
the critical points in this case are periodic orbits of the isotropic harmonic oscillator
\begin{eqnarray}
\label{eqHC7}
{\cal H}_L= E_1+E_2=2 (E_0-E_{cr}).
\end{eqnarray}
In fact, by a proper canonical rescaling of the variables in (\ref{eqHC7}), we can show that
these energy surfaces are hyperspheres and that the group generated by the
constants of motion (\ref{eqHC4}), (\ref{eqHC44}) and (\ref{eqHC6-0}) is homomorphic
to the unitary unimodular group with the topology of $S^3$\cite{ozorio,cqg}.
These energy surfaces are denoted the center manifold $S^3$ of unstable periodic orbits,
a structure that extends to the nonlinear phase apace domain about the critical points.
\par Now due to the separate conservation of $E_1$ and $E_2$ in (\ref{eqHC7}) we can show that
the center manifold in the linear neighborhood of the critical points is foliated by Clifford 2-dim
surfaces in $S^3$\cite{sommer}, namely, 2-tori $\mathcal{T}_{E_0}$ contained in the energy surface
$E_0={\rm const}$. The Clifford surfaces as well as the $S^3$ manifold containing them
depend continuously on the parameter $E_0$. These two tori will have limiting configurations
which are periodic orbits, whenever $E_1=0$ or $E_2$=0.
\par
From  equation (\ref{eqHC7}) we have that $(E_0-E_{cr})<0$ is a necessary condition for the
dynamics in the rotational sector (\ref{eqHC7}), defining a condition for the existence of the
center manifold $S^3_{E_0}$ of periodic orbits.
For $E_0=E_{cr}$ the center manifold reduces to the critical point.
By continuity as $(E_{cr}-E_0)$ increases the nonlinear extension of the center
manifold maintains the topology of $S^3$ but will no longer be decomposable into
$E_1$ and $E_2$. A detailed description of the center manifold and its nonlinear extension
will the object of the next Section.
\par The second possibility to be considered is the motion in the sector $(x,p_x)$.
In the parameter domain $(A)$ the case of $E_x$ demands a separate analysis for the two critical points,
since we have $q_x>0$ for the critical point $P_2$ so that $E_x$ corresponds to the energy associated
with the motion in the saddle sector. We remind that this is related to the fact that the
pair of eigenvalues (\ref{L1}) are real for $P_2$.
\par
We should remark that for the critical point $P_1$, in which $q_x<0$, $E_x$
is positive definite and corresponds to the rotational energy in the additional rotational
sector $(x,p_x)$ of the dynamics about $P_1$ so that the general motion about $P_1$ will have
the topology $S^1 \times S^3$. All the orbits of the dynamics about $P_1$ will
be oscillatory corresponding to perpetually nonsingular bouncing universes.
\par
In the following our focus will be in the phenomena connected to the
saddle-center-center critical point $P_2$ present in the phase space of the model.
The general dynamics in the linear neighborhood of $P_2$
is more complex and comes from the presence of the saddle sector associated with $q_x>0$,
as we discuss now.

\par
If $E_x=0$ we have two possibilities. The first is $(x=M_0,p_x=0)$
which corresponds to the motion in the center-center ($S^3$) sector already examined.
\par The second possibility is
$p_x= \pm \sqrt{24 M_0 q_x}~ (x-M_0)$ which
defines the linear stable $V_S$ and linear unstable $V_U$ manifolds of the saddle sector.
$V_S$ and $V_U$ limit regions $I$ ($E_x<0$) and regions $II$  ($E_x>0$) of motion on
hyperbolae that are solutions of the separable saddle sector $E_x=p_x^2/24 M_0 -q_x (x-M_0)^2$.
Note that the saddle sector depicts the neighborhood of $P_2$ in Figure \ref{detM}, with $V_S$ and $V_U$
tangent to the separatrices at $P_2$.
The direct product of $\mathcal{T}_{E_0} \subset S^3$ with $V_S$ and $V_U$
generates, in the linear neighborhood of $P_2$, the structure of stable ($\mathcal{T}_{E_0} \times V_S$)
and unstable ($\mathcal{T}_{E_0} \times V_U$) 3-dim surfaces that coalesce into the 2-dim tori $\mathcal{T}_{E_0}$
for times going to $+ \infty$ and $- \infty$ respectively. The energy on any orbit on these tubes
is the same as that of the orbits on the tori $\mathcal{T}_{E_0}$.
\par These 3-dim tubes are contained
in the 4-dim tubes which are respectively the product of $V_S$ and $V_U$ times the center manifold $S^3_{E_0}$ ,
in the 5-dim energy surface ${\cal{H}}=0$ (\ref{eq10091}) with $E_0-E_{cr}<0$. They
constitute a boundary for the general dynamical flow and are
defined by $E_x=0$ in the linear neighborhood of $P_2$. Depending on the sign
of $E_x$ the motion will be confined inside the 4-dim tube (for $E_x<0$) and will correspond to a
flow separated from the flow outside the tube (for $E_x>0$). The extension of structure of the 4-dim
tubes away from the neighborhood of the center manifold $S^3_{E_0}$ are now to be examined and our basic interest
will reside in the stable and unstable pair, $S^3_{E_0}\times V_S$ and $S^3_{E_0} \times V_U$, that leave the
neighborhood of $P_2$ towards the bounce.
\par
In the two following sections the nature of the center-manifolds about critical points
with a center-center sector and the typical phase space dynamics of the system are analyzed,
for the parameter domain $(A)$. The examination of some fundamental results
are extended for general cases so that they can be applied in the remaining sections of the paper.
\begin{figure*}
\includegraphics[width=8cm,height=6cm]{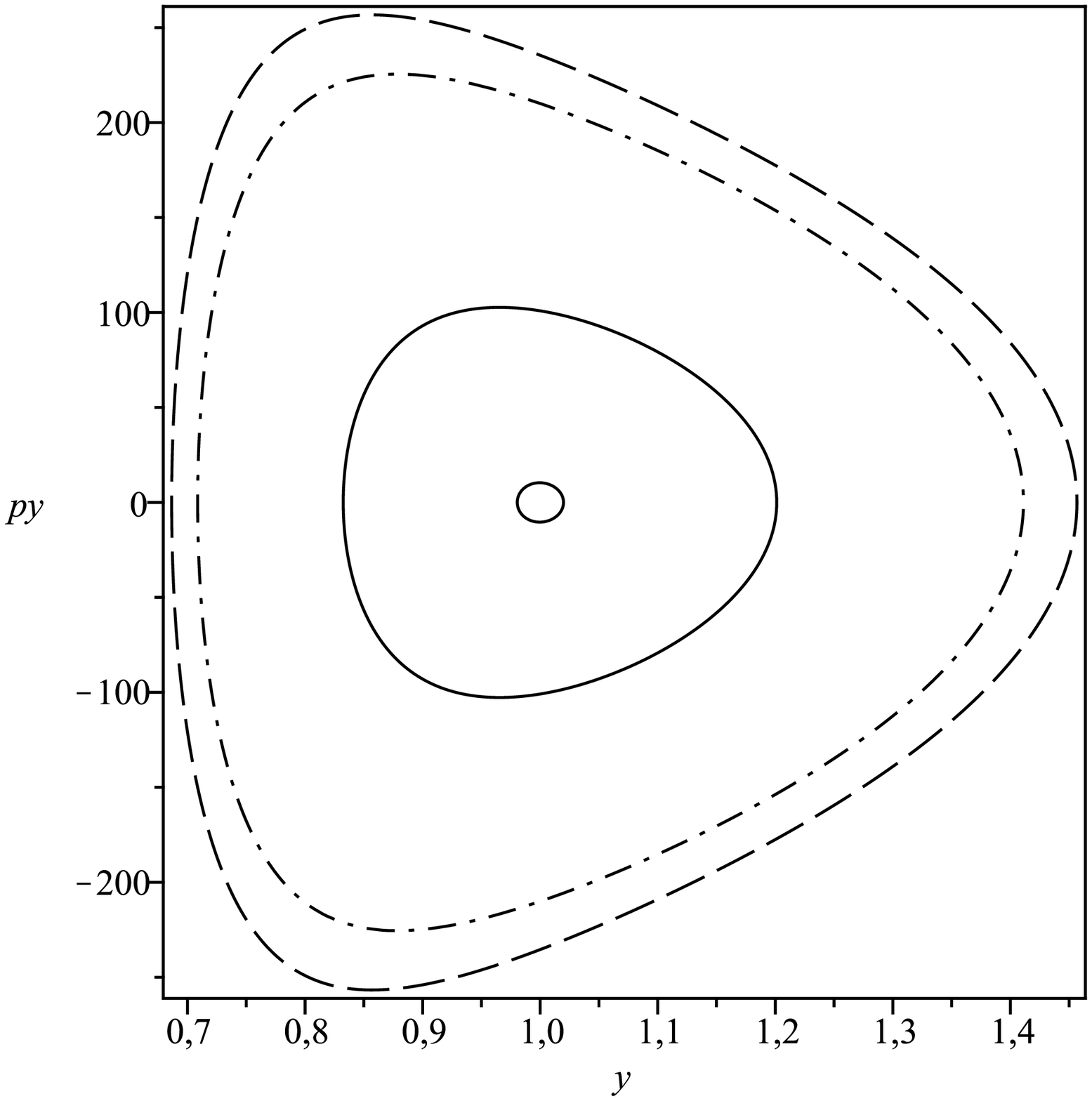}~~\includegraphics[width=8cm,height=6cm]{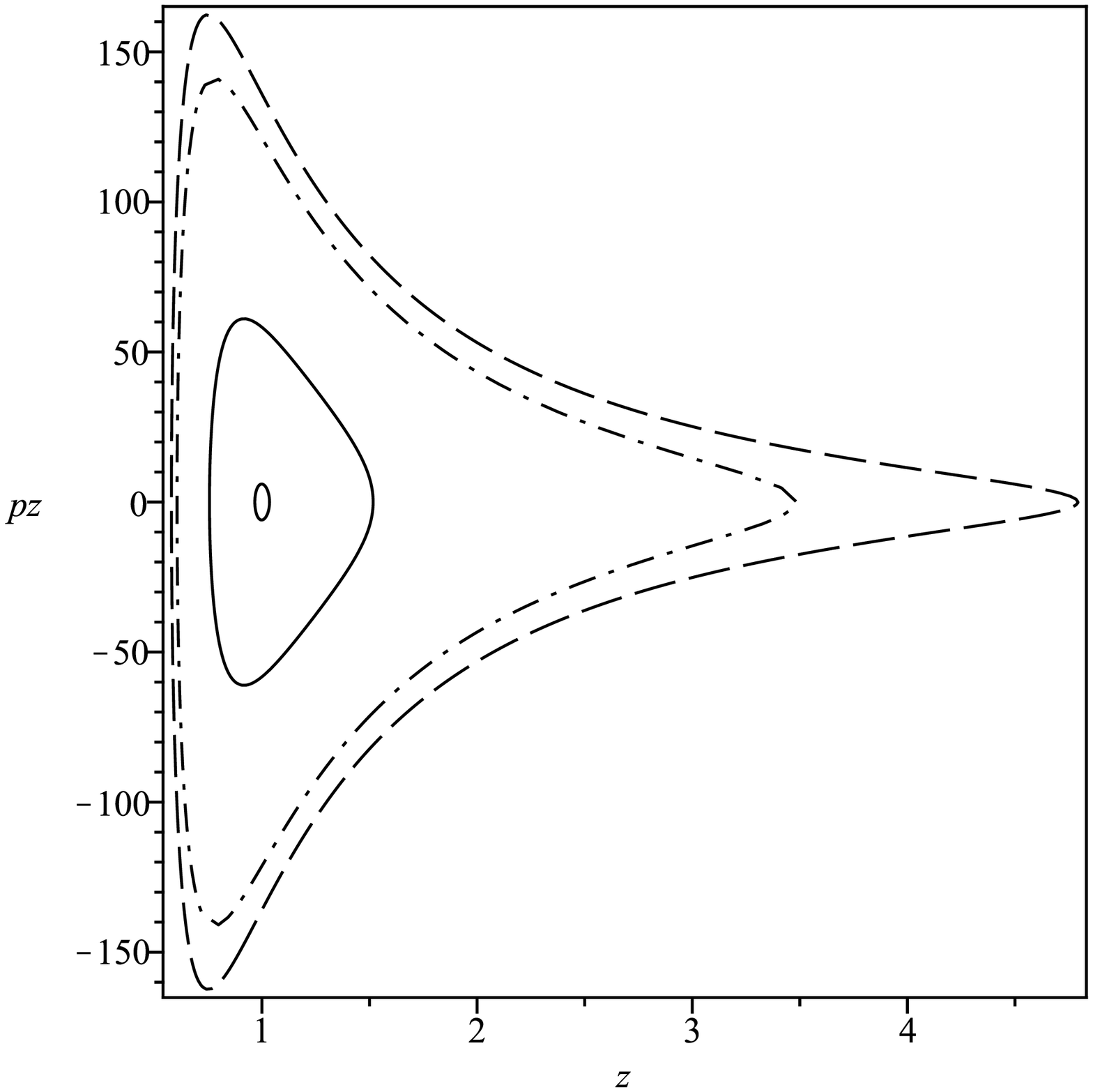}
\caption{Section $p_z=0,z=1$ (left) and section $p_y=0,y=1$ (right) of the center manifold $S^3$ (\ref{eqHcenter}), for $E_0=9.89,9.0,6.0,5.0$,
about the critical point $P_2$, with parameters (\ref{CMM1}) in $(A)$.
The energy of $P_2$ is $E_{cr_2}=9.89949505925050$. }
\label{sectionCM}
\end{figure*}
\section{The center manifold and the bouncing oscillatory dynamics\label{centermanifold}}

One of the main important uses of the canonical coordinates (\ref{eqGen})
is to give an exact analytical form of the center manifold as well as a sufficiently
accurate numerical description of the phase space dynamics in extended regions away from the critical points.
As we will see the center manifold is a fundamental structure connected with the whole oscillatory
motion in the phase space.
\par
We start by examining the nonlinear extension of the center manifold about the
saddle-center-center critical point $P_2$ restricted to the parametric domain $(A)$
(the same analysis applies to
the center manifold about the center-center-center
critical point $P_1$). In the canonical variables $(y,z,p_y,p_z)$ the equation of
the center manifold is obtained by substituting $(x=x_{cr_2},~p_x=0)$ in (\ref{eq10091}),
yielding the exact expression
\begin{eqnarray}
\label{eqHcenter}
\nonumber
{\cal H}_C&=& \frac{ y^2 p_y^2}{2 x_{cr_2}^3} + \frac{3  z^2 p_z^2}{2 x_{cr_2}^3}+\frac{x_{cr_2}}{2 z^{{4/3}}}
- \frac{x_{cr_2}}{y z^{1/3}} - \frac{x_{cr_2} y}{z^{1/3}} \\
\nonumber
&-& x_{cr_2} z^{2/3} +
\frac{x_{cr_2} z^{2/3}}{2 y^2} + \frac{1}{2} x_{cr_2} y^2 z^{2/3} + 2 x_{cr_2}^3 \Lambda \\
&+&{x_{cr_2}^3}U_{HL}(x_{cr_2},y,z)+ \frac{2 E_r}{x_{cr_2}  }+2 E_0 =0,
\end{eqnarray}
where $x_{cr_{2}}$ is the average scale factor of the saddle-center-center
critical point $P_2$. The domain of $E_0$ defining the center manifold satisfies the constraint
$E_0<E_{cr}$ as already discussed. For $E_0=E_{cr}$ the center manifold reduces to the critical point.
%
\begin{figure*}
\includegraphics[width=7cm,height=6cm]{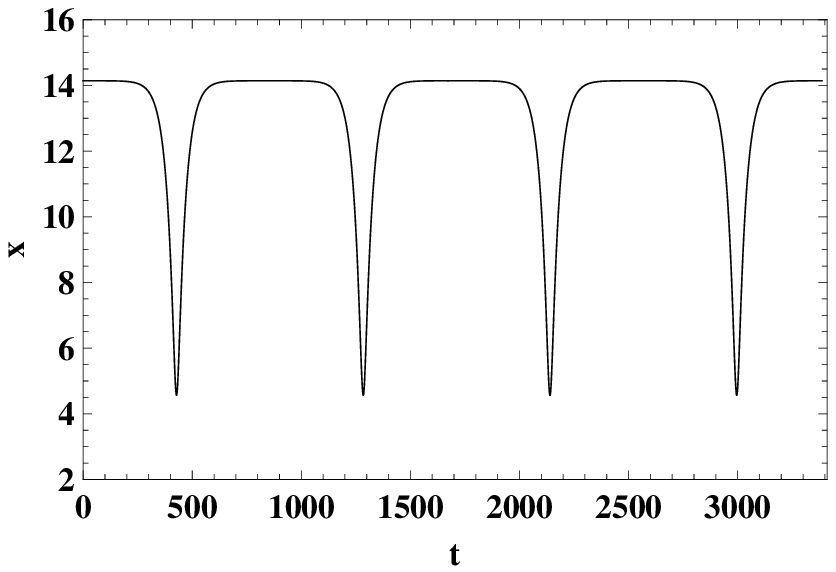}~~~~\includegraphics[width=7.5cm,height=6cm]{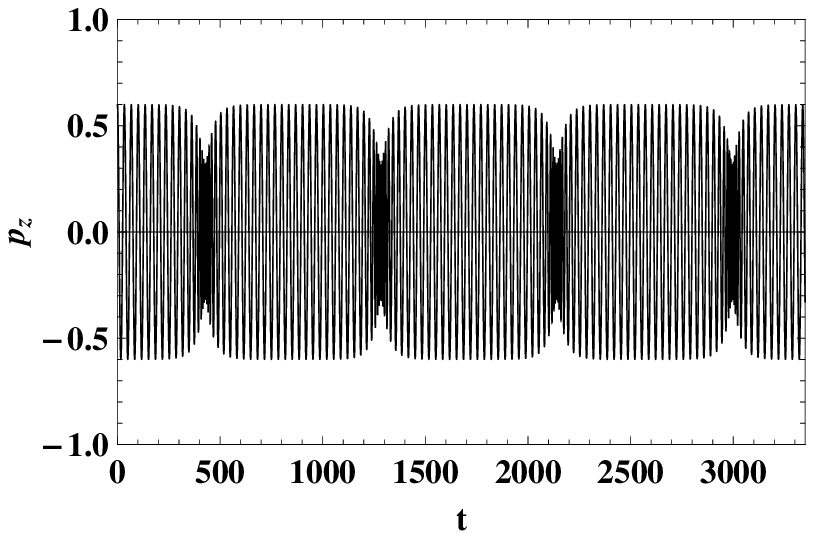}
\caption{(Left) The evolution of the scale factor $x(t)$ for a periodic orbit having infinite bounces,
with parameters (\ref{CMM1}) in $(A)$ and initial conditions on a $S^1$ section of the center manifold.
(Right) The evolution of the oscillatory mode $p_z(t)$. We note that
the frequency of the mode increases substantially as the orbit bounces.}
\label{CMsection1}
\end{figure*}
In this section we are restricted to the parameter domain $(A)$ and, for illustrative
purposes, we will initially adopt the parameters
\begin{eqnarray}
\label{CMM1}
\nonumber
\Lambda&=&0.001,~~~Er=10,\\
\alpha_{31}&=&10^{-4},~~~\alpha_{32}=10^{-5},~~~\alpha_{33}=0.00092,\\
\nonumber
\alpha_{21}&=&-100,~~~\alpha_{22}=580/3.
\end{eqnarray}
For this parameter configuration the critical point $P_2$ is given
by $(x_{cr_2}=14.14213521493478,p_x=0,y=1,p_y=0,z=1,p_z=0)$
with a corresponding critical energy $E_{cr_2}=9.89949505925050$.
The center manifold (\ref{eqHcenter}) is illustrated in Figures \ref{sectionCM} where
we display its sections $(p_z=0,z=1)$ and $(p_y=0,y=1)$
about $P_2$ for several decreasing values of $E_0=9.89,~9.0,~6.0,~5.0$,
illustrating the $S^3$ topology of (\ref{eqHcenter}) and its deformation as the parameter
$(E_{cr}-E_0)$ increases.
\par
The center manifold (\ref{eqHcenter}) is the locus of unstable periodic orbits or
of oscillatory orbits of the system and organizes the finite phase space dynamics.
In the case of the center manifold about the
saddle-center-center $P_2$, let us consider for simplicity
the section $(y=1,p_y=0)$ of Figure \ref{sectionCM} (right) for the
energy $E_0=9.8994$. This section has the topology of $S^1$ and
is a solution of the constraint (\ref{eqHcenter}), ${\cal H}_C(y=1,p_y=0,z,pz,E_0)=0$,
with $(E_{cr}-E_0)\simeq 9.5 \times 10^{-5}$. The points of this section are initial
conditions for perpetually bouncing orbits, propagated forward or backward in time, as can be
verified numerically.
Let us take for instance the point $(z=1,p_z=0.598741056178016)$ as initial conditions
on $S^1$. The result of the
dynamics is a perpetually bouncing universe illustrated in Figures \ref{CMsection1}.
In  Figure \ref{CMsection1} (left) we can see that the orbit undergoes a long time oscillation on the
center manifold -- namely at ($x=x_{cr}=14.14213521493478,p_x=0$) -- with short intervals of
escaping from this neighborhood towards the bounces at $x_b \simeq 4.565245$, with
a period between the bounces of $\Delta t \simeq 855.975$.
\begin{figure}
\includegraphics[width=8.0cm,height=5cm]{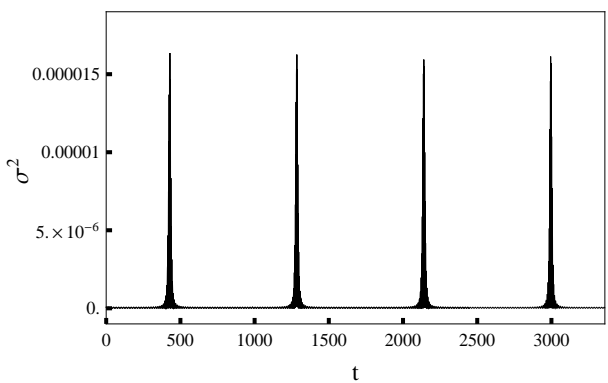}
\caption{Plot of the evolution of the anisotropy parameter $\sigma^2$
related to the periodic orbit with infinite bounces displayed in Figs. \ref{CMsection1},
showing a relative large amplification in the oscillations when
the orbit visits a neighborhood of the bounces.}
\label{shear}
\end{figure}
\par
The oscillatory behavior of the modes $(z,p_z)$ is illustrated in Figure \ref{CMsection1} (right)
showing long time oscillations about the center
manifold with short intervals in which the orbit visits the bounce and returns again to the neighborhood of the
center manifold. We note that the frequency of the mode $p_z(t)$ increases substantially at the bounces.
Analogous behavior is present in the variable $z(t)$, as expected.  These patterns were verified
for a long time evolution. Actually in all our numerical treatment the
Hamiltonian constraint (\ref{eq10091}) is conserved within a numerical error $\leq 10^{-13}$ for the whole
computational domain. These orbits constitute a set of perpetually bouncing periodic orbits present
in the dynamics of the model.
\par
Summarizing, as discussed already in the previous section, the orbits
$(x(t),p_x(t),y(t),p_y(t),z(t),p_z(t))$ emerge from the $S^3$ center manifold
towards the bounce generating the 4-dim stable and unstable cylinders $R \times S^3$, the motion
along the cylinders being obviously oscillatory. For simplicity we restricted our numerical illustration
to the motion in the invariant submanifold $(y=1,p_y=0)$ with initial conditions taken on the
1-dim manifold $S^1 \subset S^3$ defined by ${\cal H}_C(y=1,p_y=0,z,pz)=E_{cr}-E_0$
where $E_0=9.8994$, corresponding to an energy of rotational motion
in the sector $(z,p_z)$ of $\simeq 1.52 \times 10^{-4}$.
In fact the orbits discussed above are strictly periodic bouncing orbits and therefore are not orbits homoclinic
to the center manifold in which case they would take an infinite time to its return to
the center manifold.
\begin{figure}
\includegraphics[width=8cm,height=6cm]{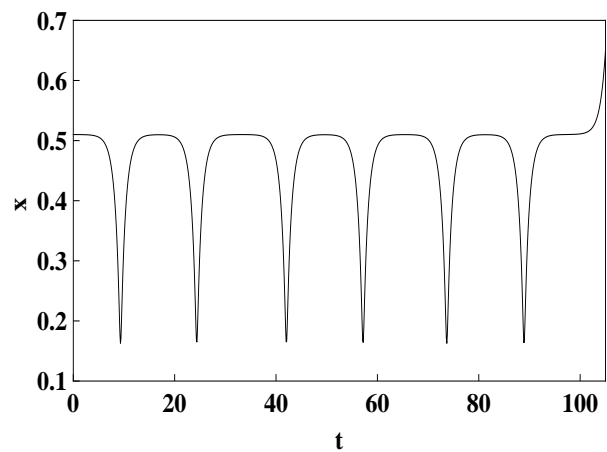}~~~~~~
\caption{The evolution of the scale factor $x(t)$ for an orbit with parameters (\ref{CMMM1}) in $(A)$ exhibiting six bounces before
the orbit escapes to the de Sitter attractor at infinity.}
\label{CMsection11}
\end{figure}
\begin{figure*}
\includegraphics[width=8cm,height=6cm]{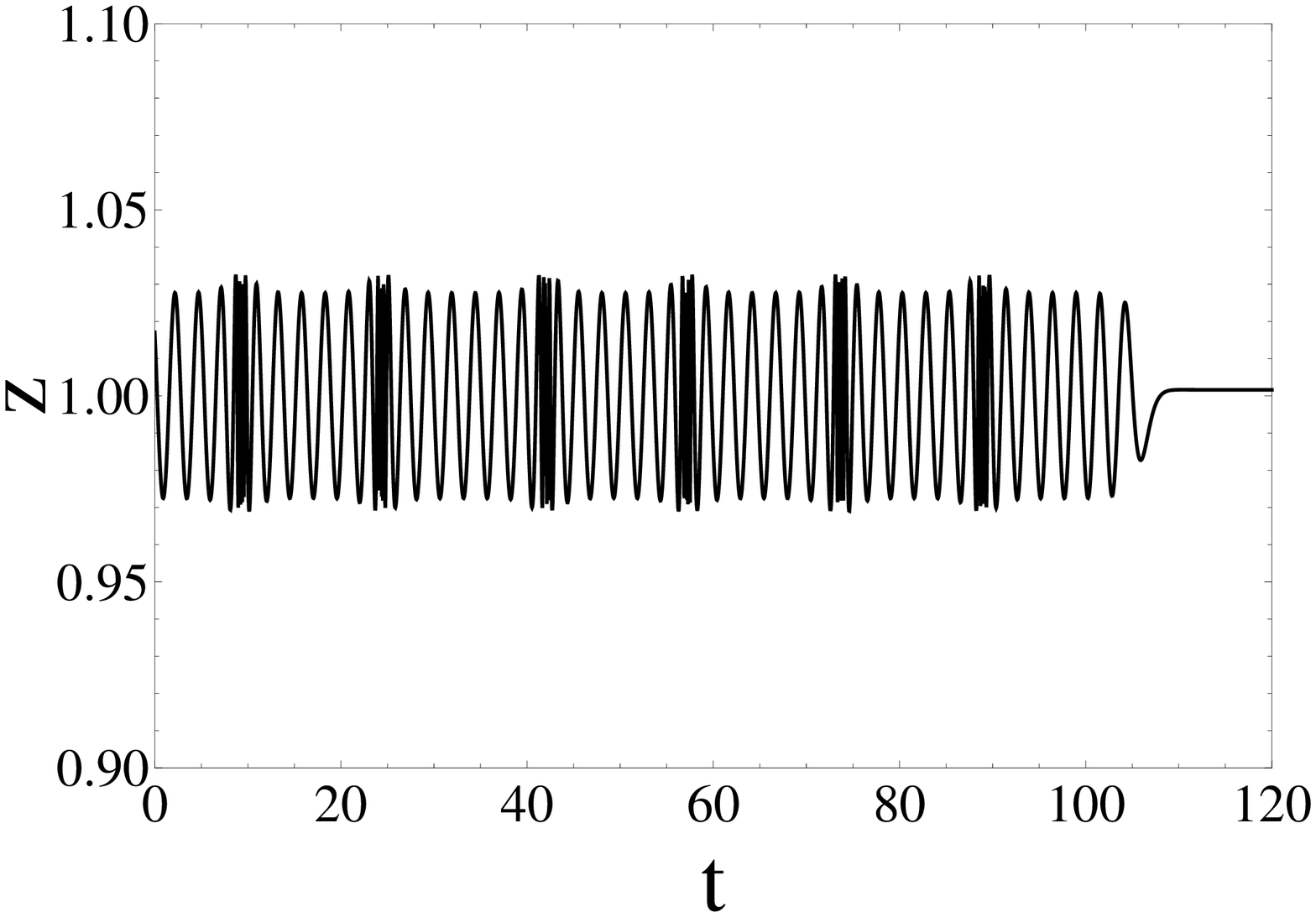}~~~~~~\includegraphics[width=7.5cm,height=6cm]{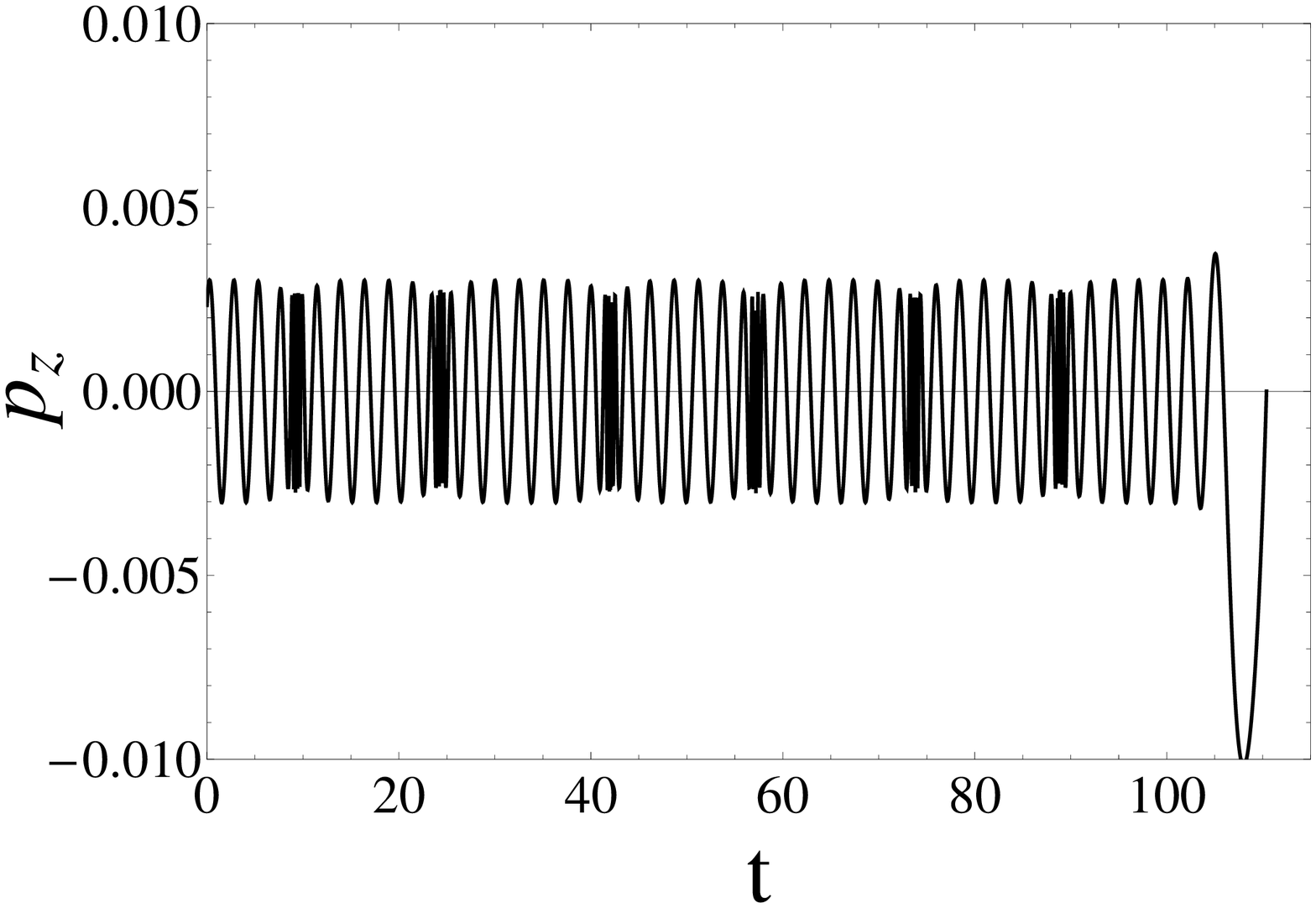}
\caption{The evolution of the oscillatory modes $p_z(t)$ (left) and $z(t)$ (right) of the orbit of Fig. \ref{CMsection11}.
The frequency of the modes increases substantially in the neighborhood of the bounces. As the orbit approaches the de Sitter
attractor the variables approach the constant values $(z\sim 1,p_z \sim 0)$ as expected.
}
\label{CMsection12}
\end{figure*}
\begin{figure*}
\includegraphics[width=8cm,height=5cm]{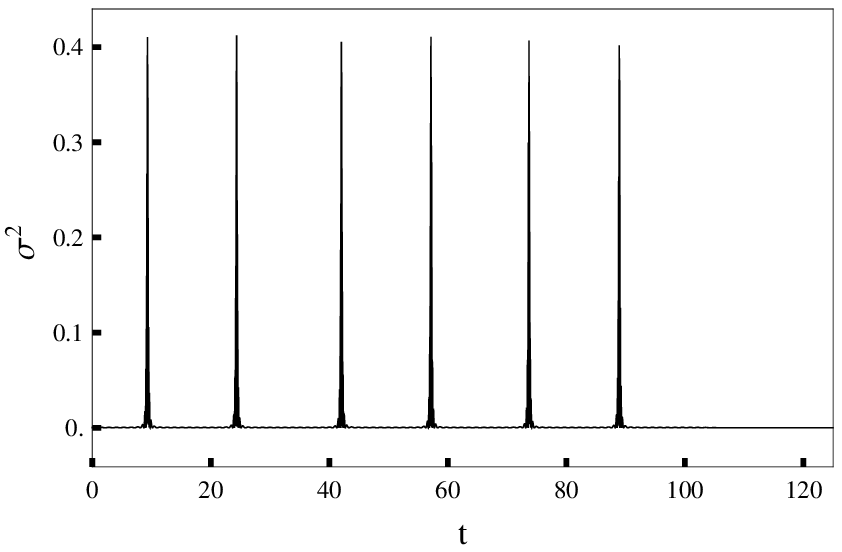}~\includegraphics[width=8cm,height=5cm]{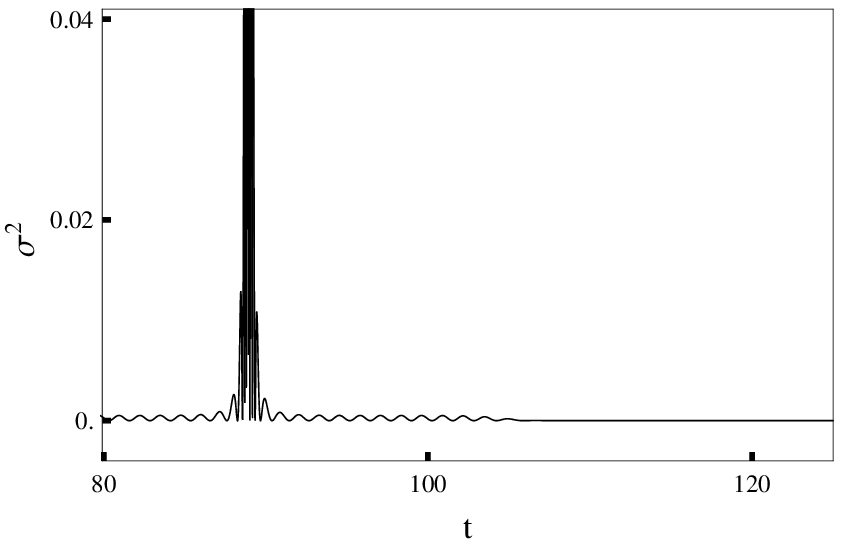}
\caption{(Left) Plot of the evolution of the shear parameter $\sigma^2$ for the orbit of Figs. \ref{CMsection11}.
(Right) Amplification of the final part of the signal showing that the shear
becomes zero as the orbit reaches the de Sitter attractor.}
\label{htMM0}
\end{figure*}
\par
Finally for future reference we will introduce a new quantity of the dynamics, the
square of the shear tensor $\sigma_{\alpha \beta}$ associated to the four velocity vector $V^{\alpha}= \delta^{\alpha}_{0}$
of a comoving observer with the matter content of the model. In the coordinate system
of the metric (\ref{ds}) for the gauge $N=1$ we obtain, after a straightforward calculation,
that
\begin{eqnarray}
\label{shear1}
\sigma^2 \equiv \frac{2}{3}~\sigma^{\alpha \beta} \sigma_{\alpha \beta}=\frac{3 (z~p_z)^2+(y~ p_y)^2}{3x^6},
\end{eqnarray}
in the new canonical variables (\ref{eqGen})-(\ref{eqGen2}).
We can see that $\sigma^2$ -- which is a measure of the anisotropy of the motion --
is basically associated with the rotational modes of the
system and has a smooth behavior. This is illustrated in Fig. \ref{shear} where we display $\sigma^2$ versus $t$
for the perpetually bouncing orbit of Figs. \ref{CMsection1} in the parameter domain $(A)$,
showing a relatively large amplification as the orbit visits a neighborhood of the
bounces.
\par
On the other hand the parameter $\sigma^2$ can play a role in the
recognition and characterization of patterns in the phase space dynamics, connected to
the presence of a saddle-saddle-saddle critical point. It will constitute an
important numerical indicator of the existence of highly anisotropic momentum attractors
in the parameter domain $(B)$, as discussed later, where the dynamics is highly unstable
due to the presence of a saddle of multiplicity two.
\par
To complete the present section we now examine a new set of parameters in $(A)$, for which a
distinct dynamical pattern connected to the saddle-center-center $P_2$ is present ,
namely, the presence of oscillatory orbits that escape to the de Sitter attractor
at infinity after a finite number of bounces. The parameters are
\begin{eqnarray}
\label{CMMM1}
\nonumber
\Lambda&=&1,~~~E_r=0.1,\\
\alpha_{31}&=&0.002,~~~\alpha_{32}=0,~~~\alpha_{33}=-0.013,\\
\nonumber
\alpha_{21}&=&0,~~~\alpha_{22}=-13/60.
\end{eqnarray}
For this parameter configuration the saddle-center-center $P_2$ has coordinates
$(x_{cr_2}=0.51007113736321,p_x=0,y=1,p_y=0,z=1,p_z=0)$,
with the corresponding critical energy $E_{cr_2}=0.22015171926053$. The initial conditions
of the orbits are taken on the $S^1$ section $(y=1,p_y=0)$ of the center manifold $S^3$ about $P_2$,
defined by the constraint ${\cal H}_C(y=1,p_y=0,z,pz,E_0)=0$,
for the energy $E_0=0.2201$.
We take, for instance, $(z=1.017048758412991,p_z=0.0023417842316)$. We evolve this initial condition
forward in time, along a neighborhood of the unstable cylinder emanating from the center manifold
towards the bounce as illustrated in Figures \ref{CMsection11}. The evolution of the
scale factor $x(t)$ is displayed in Fig. \ref{CMsection11}
where we see that the orbit undergoes $6$ bounces before escape to the
de Sitter attractor at infinity.
We should note that, contrary to the set of perpetually bouncing periodic orbits examined
previously, these orbits are non-periodic but oscillatory, since
the values of the coordinate $x_b$ of the bounces actually vary between $\simeq (0.162626,~0.163973)$,
with time intervals between the $6$ bounces being respectively $\Delta=[15.0,~ 17.7,~15.2,~16.5,~ 15.2]$.
The evolution of the oscillatory modes $p_z(t)$ (left) and $z(t)$ of the orbit of Fig. \ref{CMsection11}
is shown in Figs. \ref{CMsection12}. The frequency of the oscillatory modes increases substantially at the bounces.
As the orbit approaches the de Sitter attractor the variables approach the constant
values $(z\sim 1,p_z \sim 0)$ as expected.
In fact we must remark that in our numerical evaluations the Hamiltonian constraint (\ref{eq10091})
is conserved, with a numerical error $\leq 10^{-13}$ for the whole
computational domain. In the case of the $6$-bounces orbit of Figs. \ref{CMsection11}-\ref{CMsection12}
the Hamiltonian constraint is violated when $t \simeq 110.4$, when we stop computation.
At this time $p_z=0$ and $z \simeq 1$, with $x$ sufficiently large and $p_x \simeq 0$.
\par The evolution of the anisotropy parameter for these orbits is illustrated in Figs. \ref{htMM0}
where we plot $\sigma^2$ for the whole time domain until the orbit
reaches the de Sitter attractor. The figure at the bottom amplifies the final part of the
signal showing that the shear is zero at the de Sitter attractor, as should be expected.
\par Finally it is worth remarking that the $6$-bounces orbits of Figs. \ref{CMsection11}-\ref{CMsection12},
when propagated backward in time from its initial conditions (namely, about a neighborhood
of the stable cylinder) towards the bounce, would undergo just one bounce before escaping to the
de Sitter attractor at infinity.
\par The sets of orbits discussed in the present section characterize the oscillatory
and periodic modes present in the phase space dynamics of the system.  We should mention that
they can be related to results of Misonoh, Maeda and Kobayashi\cite{maeda} modulo their
use of the non-canonical variables $(a, \beta_{+},\beta_{-})$, and of
a distinct parametrization, where
\begin{eqnarray}
a=2x,~~ \beta_{+}=(\ln z) /6,~~\beta_{-}=\sqrt{3}~(\ln y)/6.
\end{eqnarray}
In these variables the shear parameter is expressed
\begin{eqnarray}
\label{shear2}
\nonumber
\sigma^2 = 4 ~({{\dot {\beta}}_{+}}^2 + {{\dot {\beta}}_{-}}^2).
\end{eqnarray}
We remark that, without loss of generality and for numerical simplicity, the dynamics
was restricted to the invariant submanifold $(y=1,p_y=0)$.
\par Now we are led to examine the nonlinear extension of the stable and unstable
2-dim cylinders $R\times S^1$. In order to realize this construction numerically
we do not make use here of the displacing (in the direction of the unstable cylinder) of initial conditions taken
on the invariant center-manifold, as the shooting method in \cite{wiggins1}, but instead we make use
of the instability of the motion on the center manifold which computationally conserves the Hamiltonian
constraint (\ref{eq10091}) for all $t$, within and error $\leq 10^{-13}$.
\par
\begin{figure}
\includegraphics[width=6cm,height=6cm]{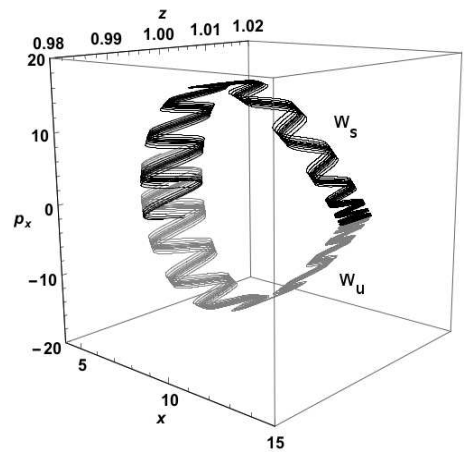}
\caption{A numerical illustration of the unstable cylinder (gray)  and the stable cylinder (black),
spanned both by $10$ orbits with initial conditions taken on a circle in the domain $(z, p_z)$ of the center
manifold for $E_0 = 9.8994$, emerging towards the bounce. The parameter configuration is given in (\ref{CMM1}).
This numerical simulation was implemented for a time domain corresponding to just one bounce,
with initial conditions $(x_0=14.142135621521241$, $p_{x_0}=0$, $y_0=1$, $p_{y_0}=0$), the $(z,p_z)$
coordinates being a solution of the constraint (\ref{eqHcenter}).}
\label{htMM}
\end{figure}

 In Fig. \ref{htMM} we illustrate the unstable $W_U$ (gray) and stable $W_S$ (black) cylinders,
spanned each by $26$ orbits, emerging from the
the center manifold section $S^1$ towards the bounce, for a time domain corresponding to just one bounce,
so that both cylinders cross just once the surface of section $\Sigma: (x=x_b,p_x=0)$
where $x_b$ is the scale factor of the bounces.
This numerical simulation was implemented for initial conditions
$(x_0=x_{cr_2}=14.142135621521241$, $p_{x_0}=0$, $y_0=1$, $p_{y_0}=0$) taken on $S^1$, with
the  $(z,p_z)$ coordinates being a solution of the Hamiltonian constraint (\ref{eqHcenter}),
for the parameters (\ref{CMM1}).
Actually $W_S$ and unstable $W_U$ cylinders, emerging from the center manifold defined for $E_0=9.8994$, is a nonlinear
extension of ($\mathcal{T}_{E_0} \times V_S$) and  ($\mathcal{T}_{E_0} \times V_U$), with $\mathcal{T}_{E_0}\subset S^3$
defined in a linear neighborhood of $P_2$. These cylinders are actually composed of orbits that
have the same energy $(E_{cr_2}-E_0) \sim 10^{-5}$ of the center manifold and coalesce to it as $t \rightarrow \pm \infty$.
\par $W_S$ and $W_U$ emanate from the center manifold towards the bounce at $x=x_b$ and are
guided by the separatrix dividing the regions
$I$ and $II$ of the invariant plane (cf. Fig. \ref{detM}). We emphasize that, in the
domain $(A)$, the separatrix $(x(t),p_x(t))$ guiding the cylinders is actually a
structure inside the cylinders about which rotational motion of the two
degrees of freedom $(y,p_y)$ and $(z,p_z)$ take place\cite{ozorio,cqg}.
It is worth noticing that the projection of the
figure in the plane $(x, p_x)$ {\it shadows} the separatrix homoclinic to $P_2$
in the invariant plane. These facts will be crucial in the
characterization of the regular/non-chaotic dynamics of the system
as discussed in the next section.

\section{On Regularity and Chaos in the Invariant Submanifolds\label{chaosINV}}

As we have seen in the previous section the stable $W_S$ and unstable $W_U$ cylinders
emerging from the center manifold about $P_2$ are 4-dim surfaces
so that they separate the 5-dim energy surface defined by the Hamiltonian constraint (\ref{eq10091})
in two dynamically disconnected pieces, a fact that is fundamental in characterization of either chaos or
regular motion in the system. The occurrence of the transversal crossing of the stable cylinder
and the unstable cylinder in the neighborhood of the bounce would constitute a topological characterization
of chaos in the dynamics, a phenomenon that eventually leads to the formation of the Poincar\'e's homoclinic
tangles\cite{holmes}.
\par
Let us consider the first transverse intersection of the cylinders: a part of the flow inside
of the unstable cylinder will enter in the interior of the stable cylinder and will
be forever separated from the part of the flow that remains outside the stable cylinder.
The part that remained inside the stable cylinder will proceed along the stable cylinder
towards the center manifold about $P_2$ from where it will reenter the unstable cylinder
and proceeds eventually to a second bounce; by a new intersection a part of these orbits
will again enter the stable cylinder and proceeds back towards the neighborhood of
the center manifold, and so on. The other part of the flow outside
stable cylinder will return to the neighborhood of the center manifold and there either
escapes towards the de Sitter attractor at infinity or returns again towards the bounce
outside the unstable cylinder. The recurrence of this process constitutes an invariant
characterization of chaos in the dynamics of the system,
generating horseshoe structures
that appear in Poincar\'e maps of the system (with surface of sections $\Sigma$ taken
at the bounce $x=x_b,p_x=0$), cf. for instance \cite{ozorio,cqg}.
If we consider the transversal crossing in a section,
say at the bounce $(x=x_b,p_x=0)$, it is not difficult to see that the intersection
is a $S^2$. Therefore the intersection manifold will be a 3-dim tube of flow (with the
topology $R \times S^2$) which is contained both in the 4-dim stable cylinder and in the
4-dim unstable cylinder, and {\it homoclinic} to the center-center manifold $S^3$.
We must recall that the cylinders are actually composed of orbits that have the same energy
$(E_{cr}-E_0)$ of the center-manifold and coalesce to it as $t \rightarrow \pm \infty$.
\begin{figure*}
\includegraphics[width=8cm,height=5cm]{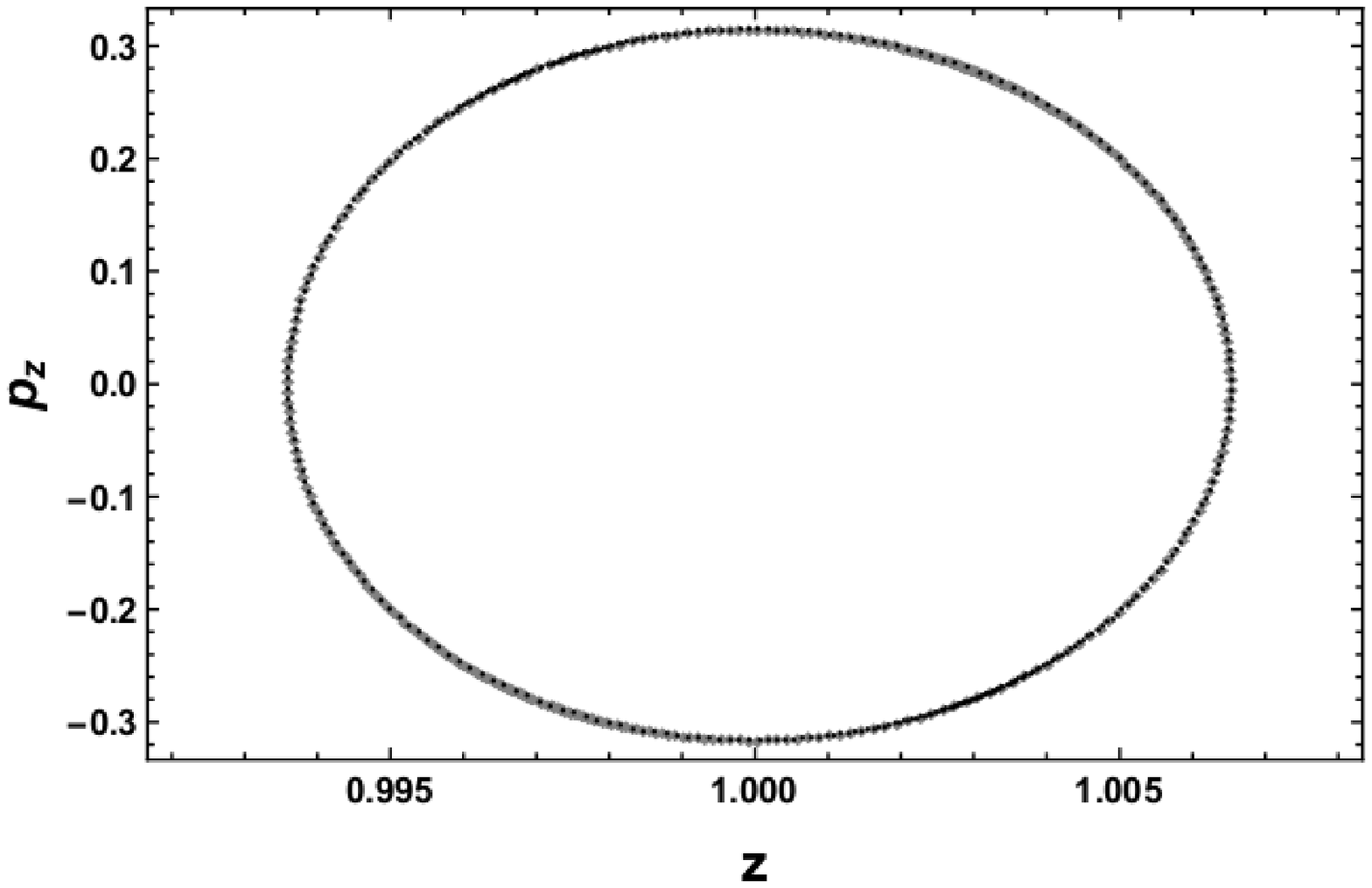}~~~~\includegraphics[width=8cm,height=5cm]{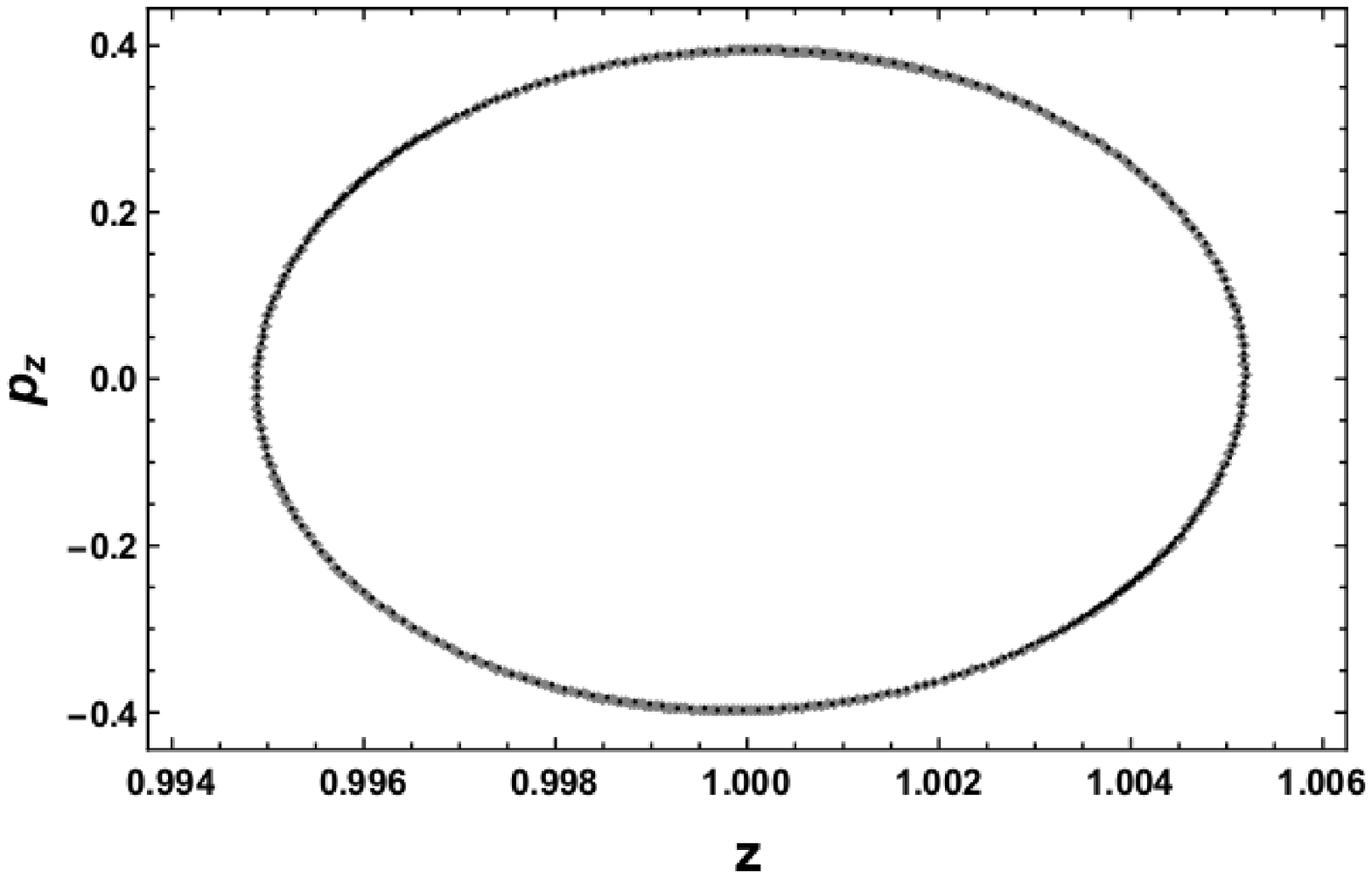}
\caption{(color on line) Poincar\'e maps of the first coalescence of the stable cylinder (black dots) and unstable
cylinder (gray diamonds) in the surface of section $\Sigma_b:(x=x_b,p_x=0)$ at the first bounce $x_b=4.565245$ (left),
and in the surface of section $\Sigma_2=(x=7,p_x=15.413)$ (right) shown in the plane $(z, p_z)$, for $E_0 = 9.8994$.
Both cylinders were spanned by $346$ orbits, with initial conditions taken on a circle in the domain $(z, p_z)$ of the center
manifold. Initial conditions and parameters are the same as in Fig. \ref{htMM}.}
\label{pm1}
\end{figure*}
\begin{figure*}
\includegraphics[width=8cm,height=5cm]{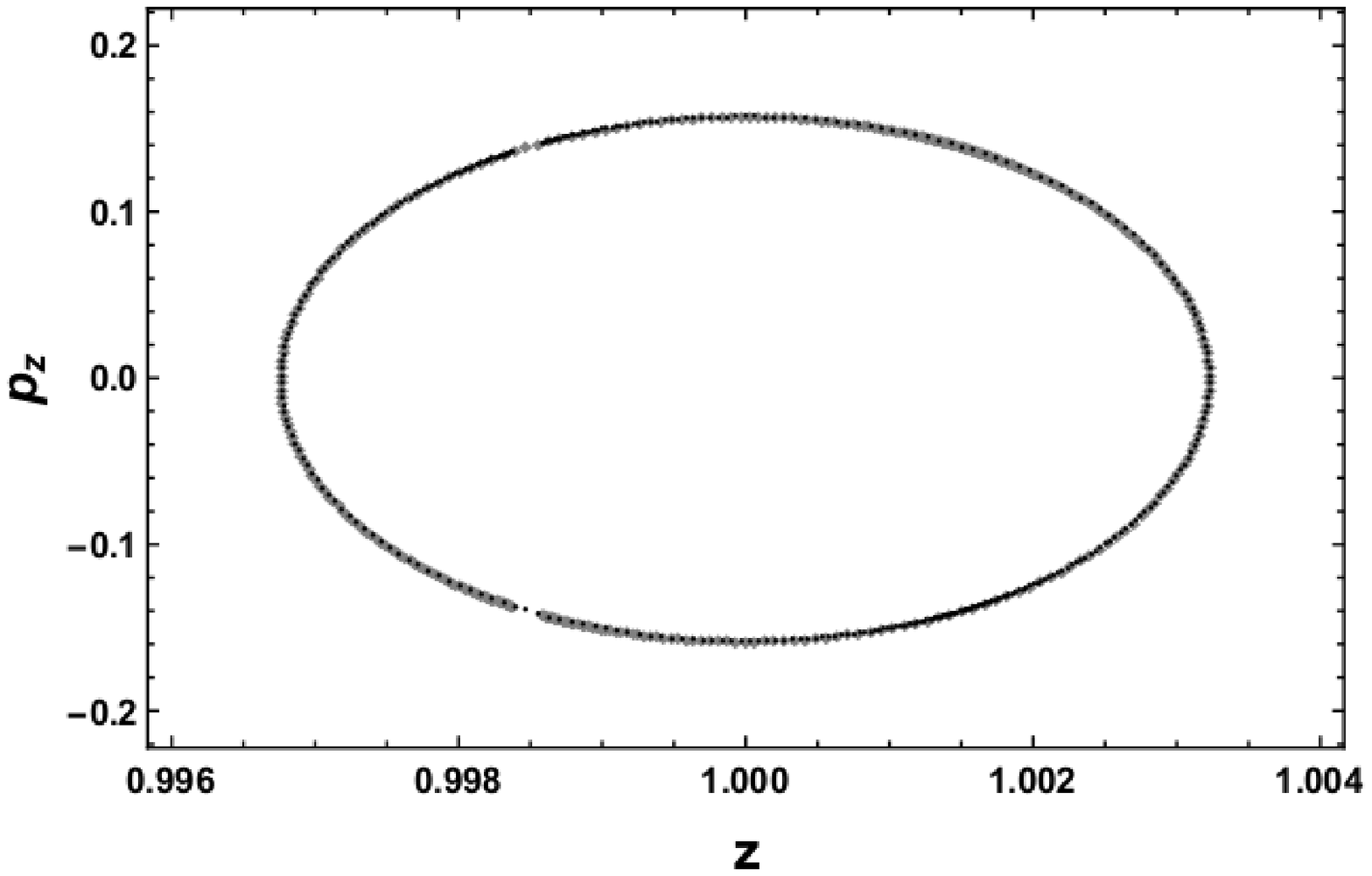}~~\includegraphics[width=8cm,height=5cm]{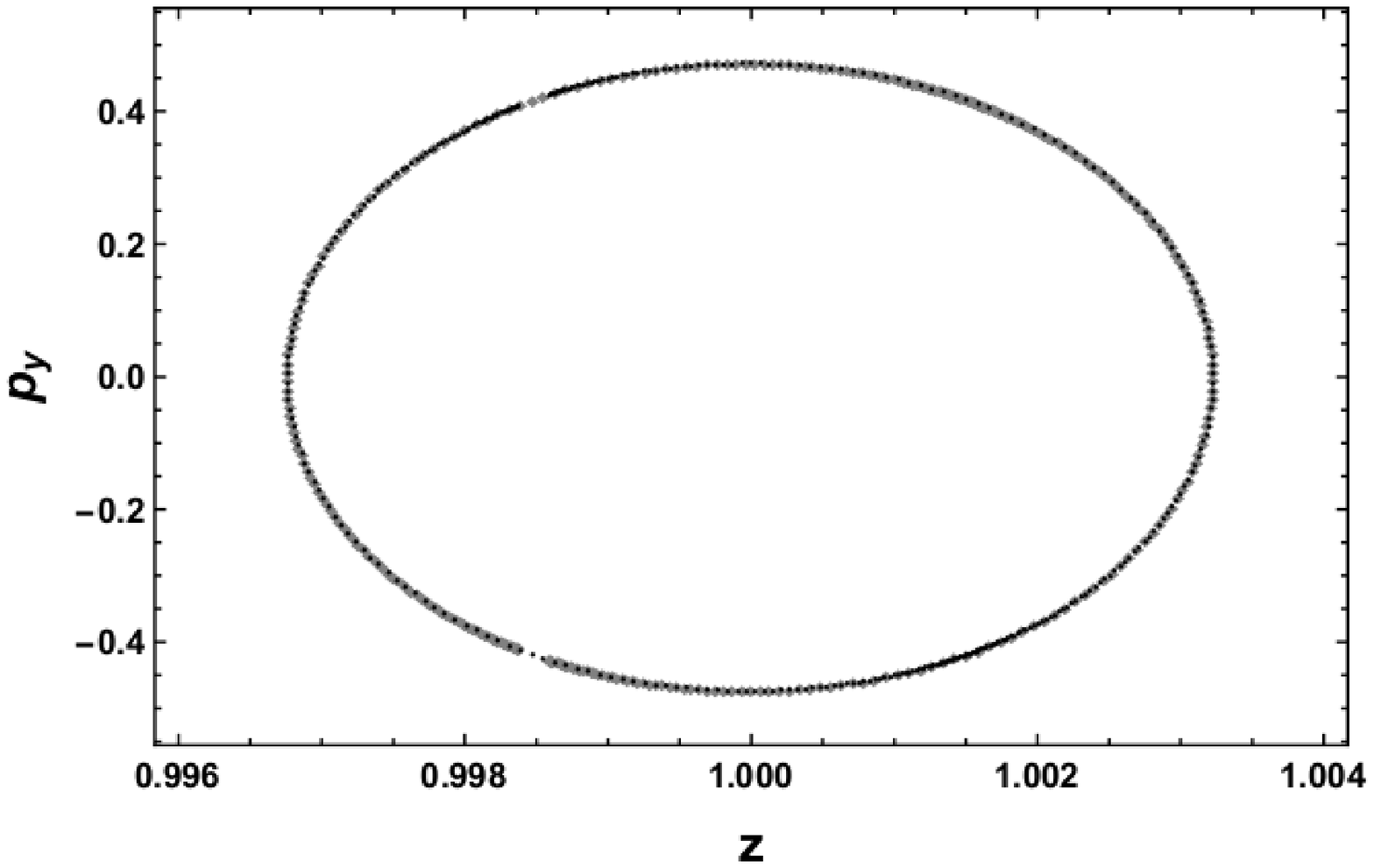}
\includegraphics[width=8cm,height=5cm]{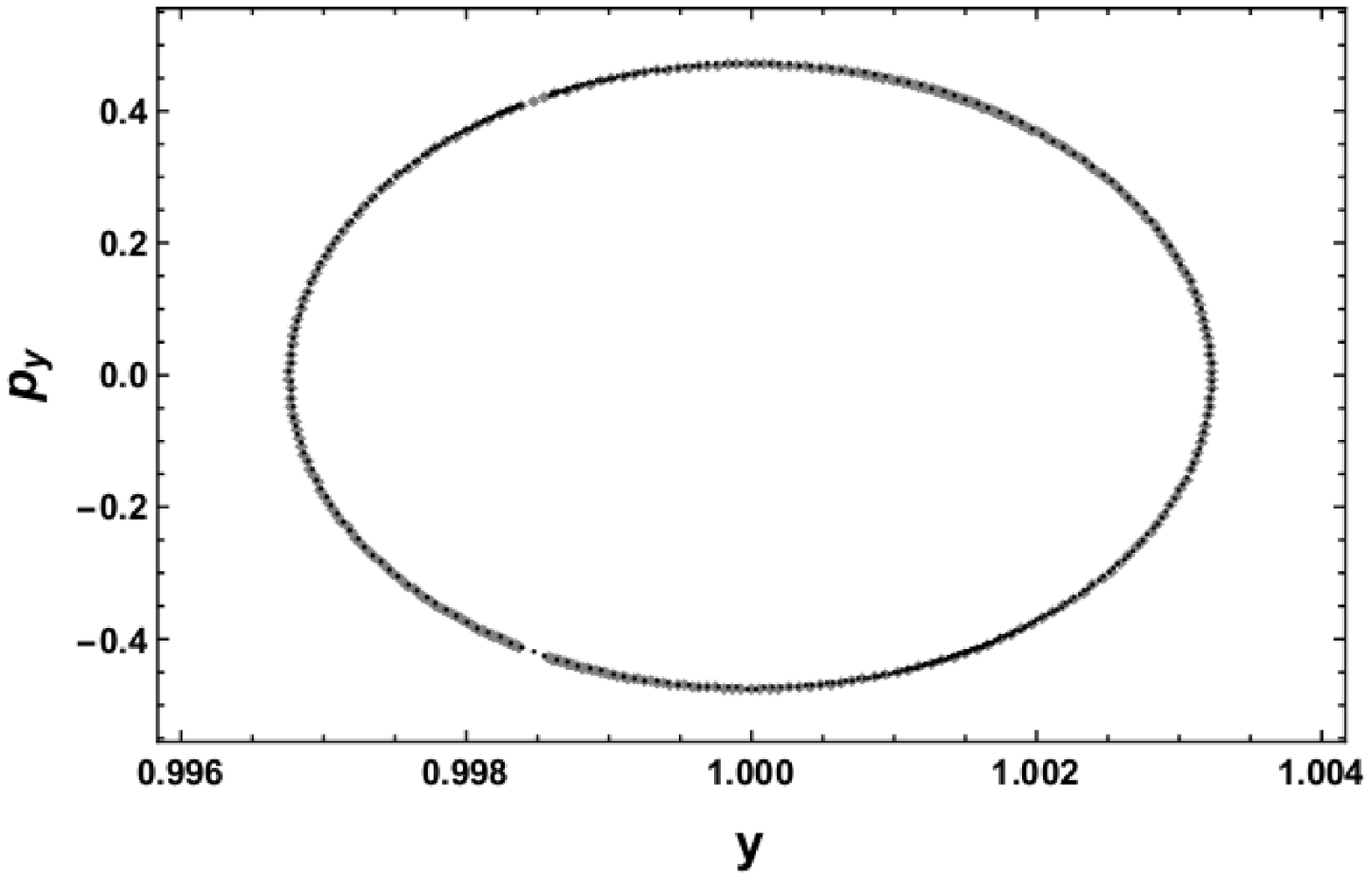}~~\includegraphics[width=8cm,height=5cm]{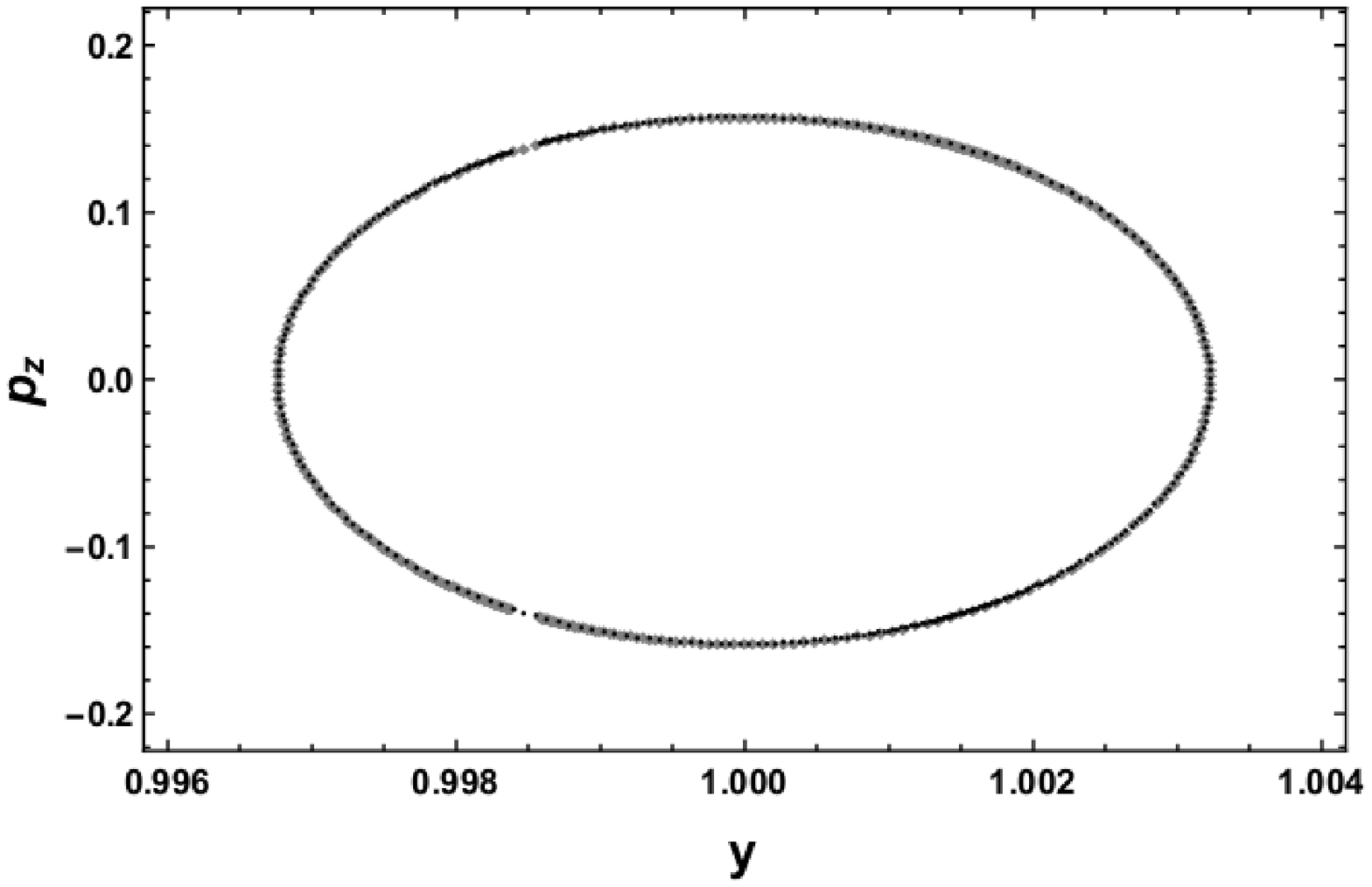}
\caption{(color on line) Poincar\'e map of the first coalescence of the stable cylinder (black dots) and unstable
cylinder (gray diamonds) in the surface of section $\Sigma_1=(x=x_b,p_x=0)$ (at the first bounce $x_b=4.56523$)
shown in the plane $(z, p_z)$ (1st panel), $(z, p_y)$ (2nd panel), $(y, p_y)$ (3rd panel) and $(y, p_z)$
(4th panel) for $E_0 = 9.8994$. Both cylinders were spanned by $173$
orbits, with initial conditions taken on a circle in the domain $(z, p_z)$ of the center
manifold. Here we fixed $\Lambda=0.001$, $E_r=10$, $\alpha_{21}=-100$, $A_2=-80$, $\alpha_{31}=10^{-4}$,
$\alpha_{32}=10^{-5}$, $A_3=10^{-5}$ so that $E_{crit} = 9.899494937274579$.
$IS_1$ initial conditions: $x=14.142135621521241$, $p_{x_0}=0$, $y_0=z_0$, $p_{y_0}=3p_{z_0}$.}
\label{ht2}
\end{figure*}
\begin{figure*}
\includegraphics[width=8cm,height=5cm]{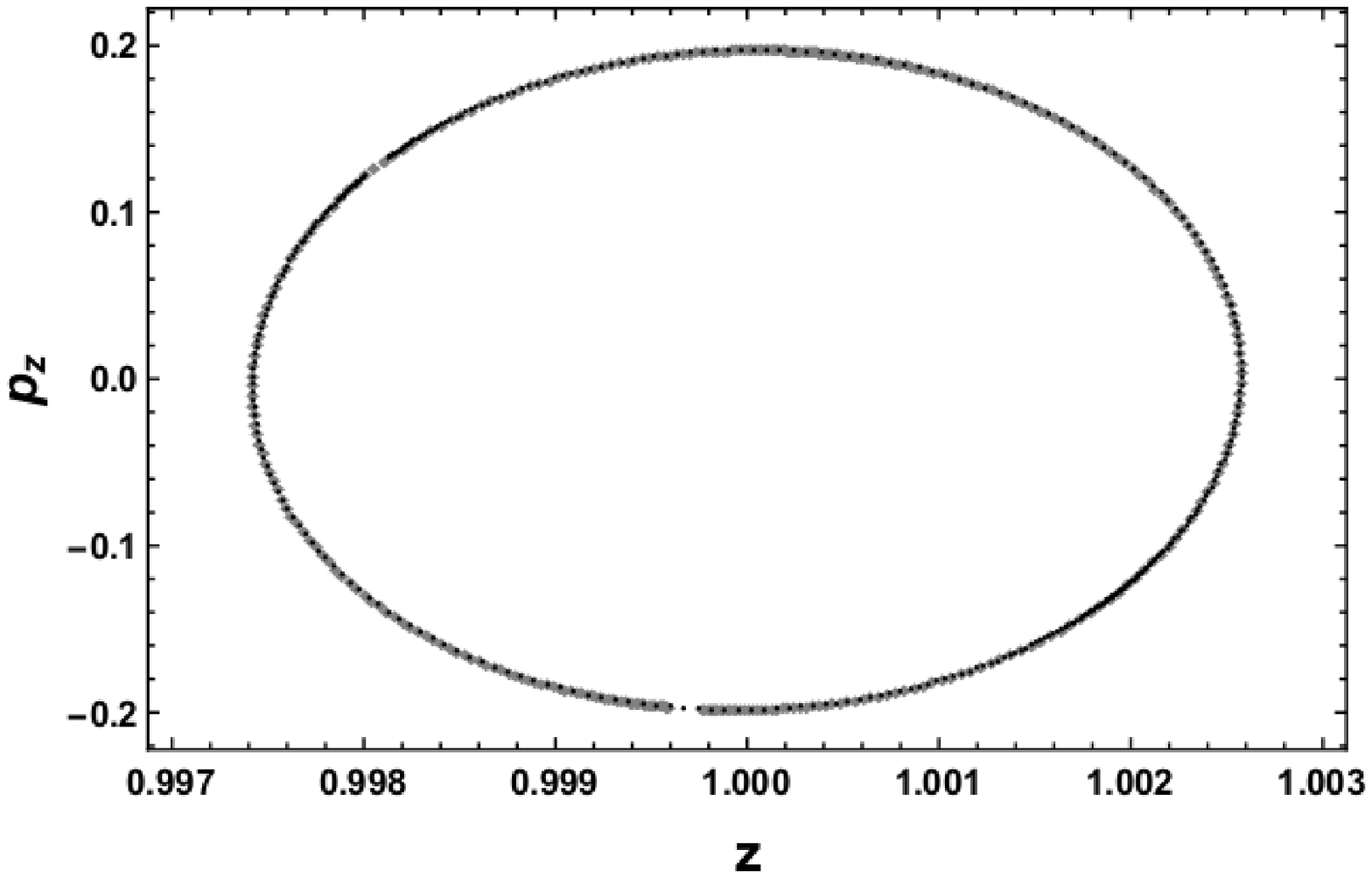}~~\includegraphics[width=8cm,height=5cm]{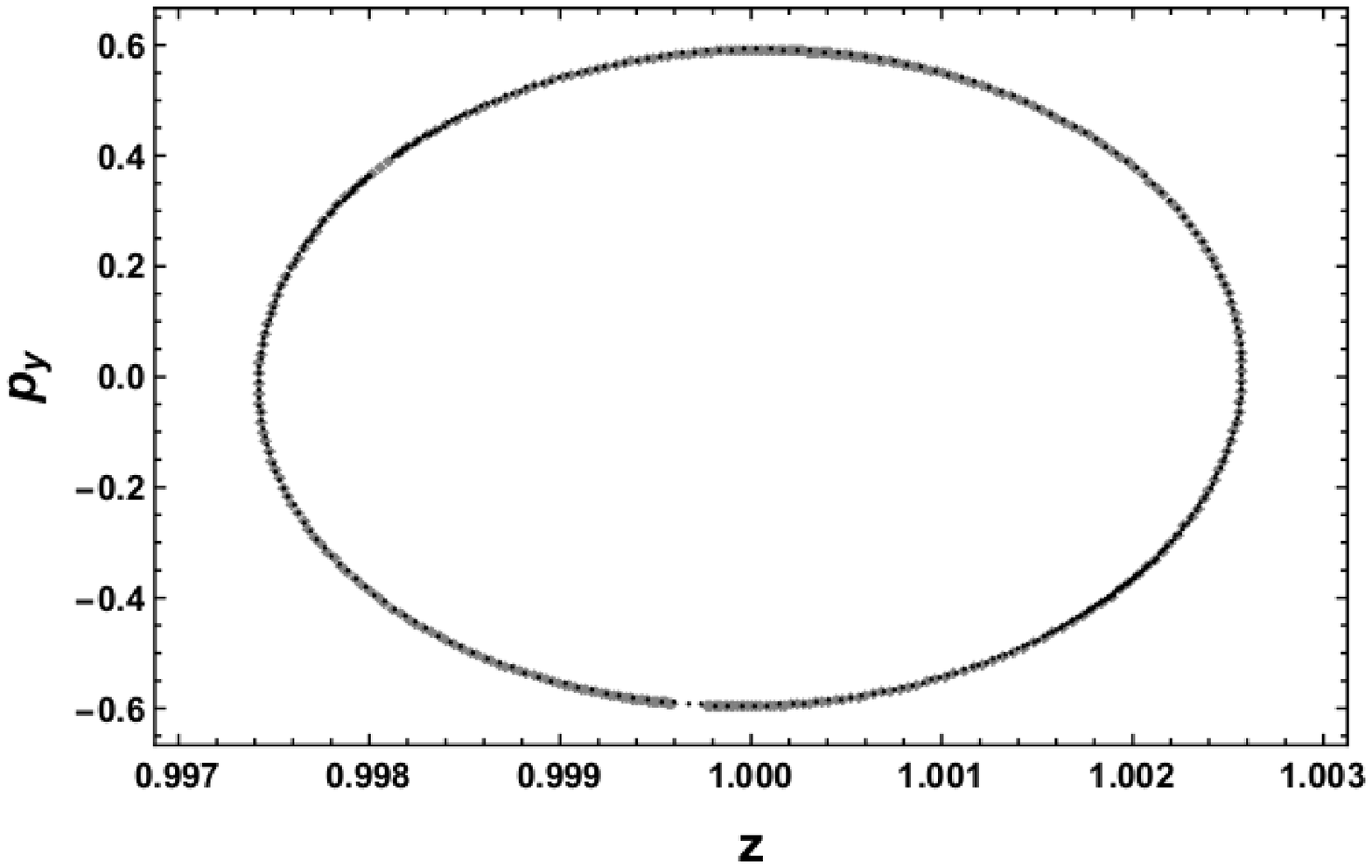}
\includegraphics[width=8cm,height=5cm]{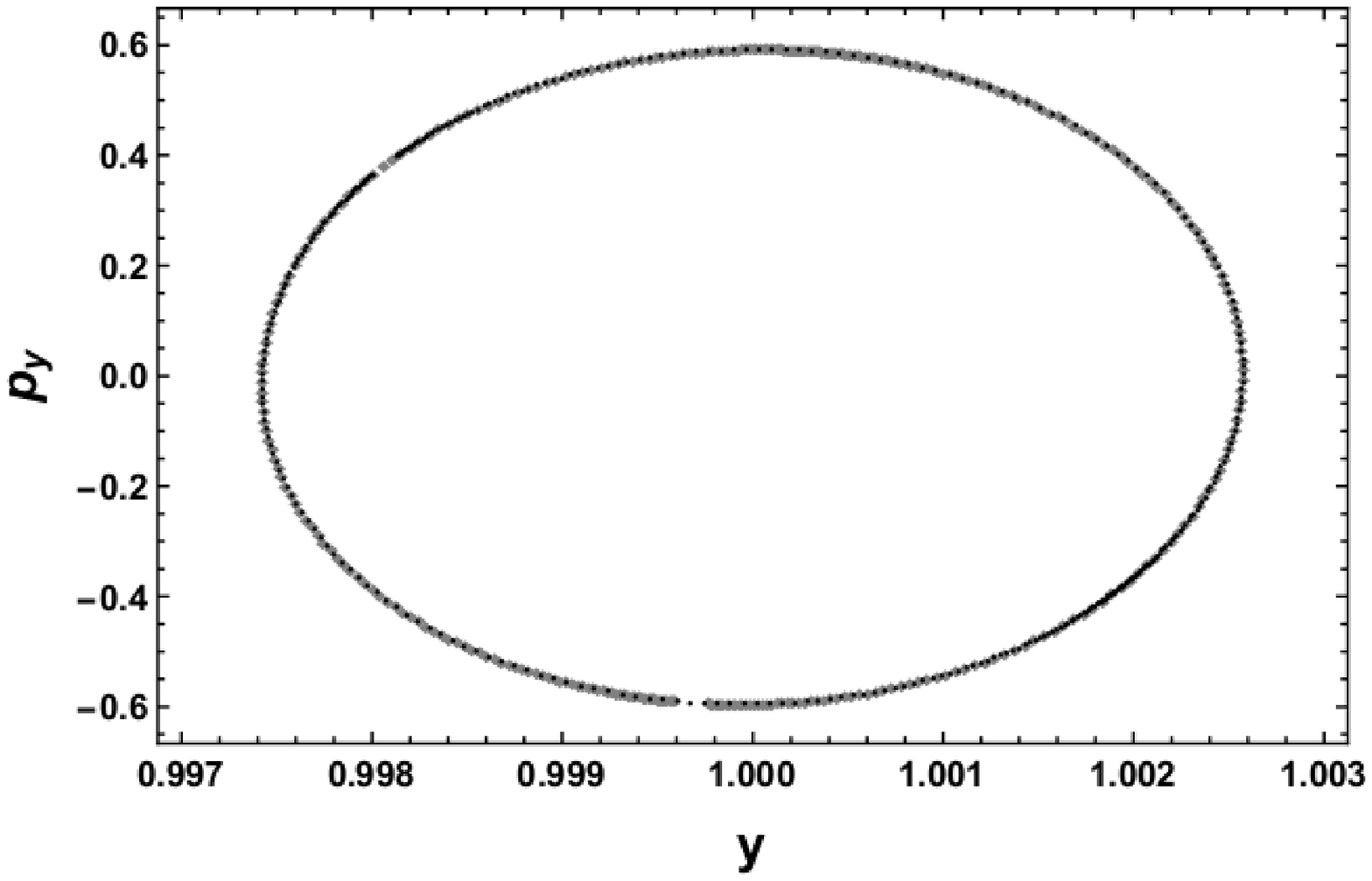}~~\includegraphics[width=8cm,height=5cm]{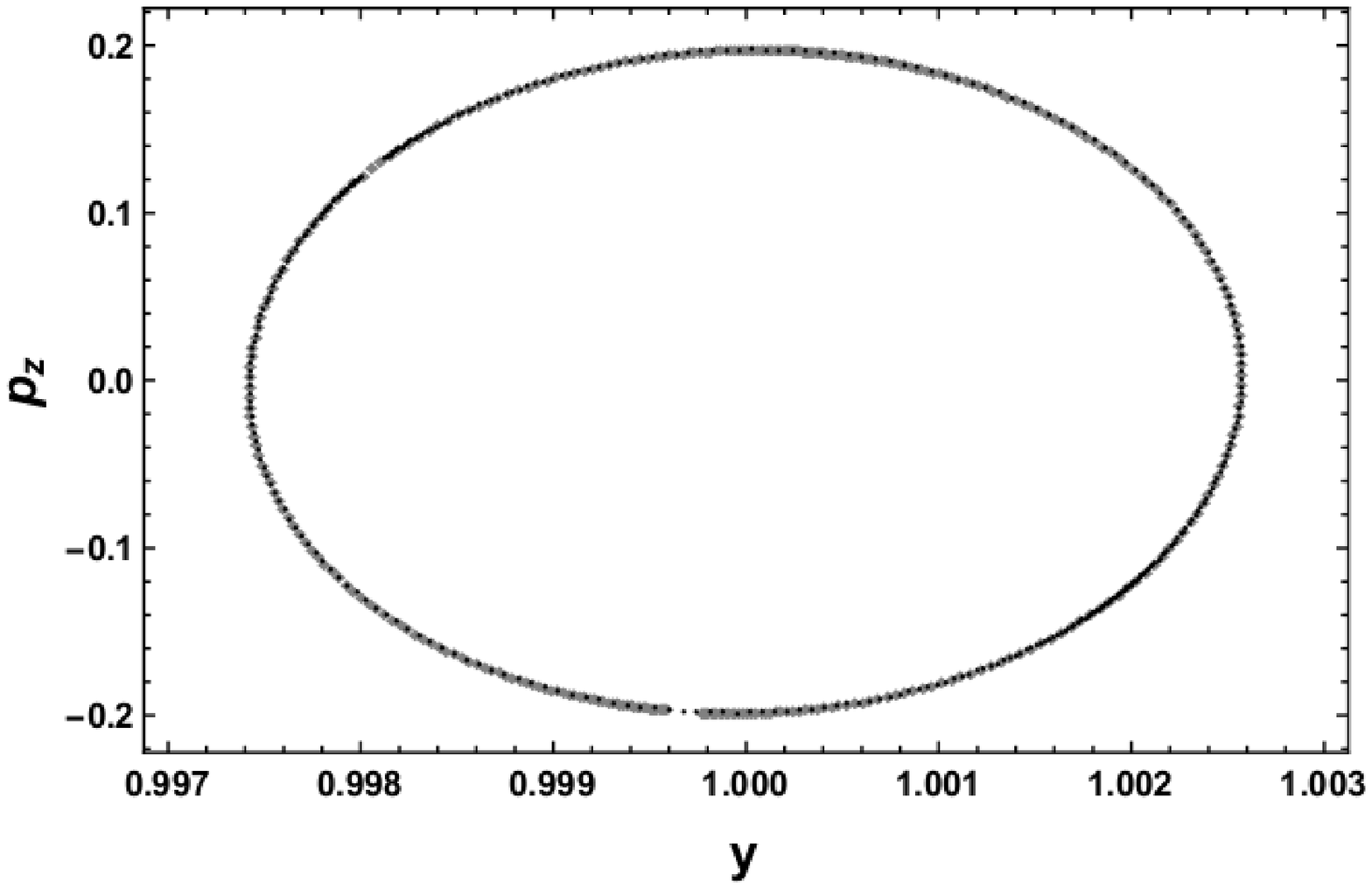}
\caption{(color on line) Poincar\'e map of the first coalescence of the stable cylinder (black dots) and unstable
cylinder (gray diamonds) in the surface of section $\Sigma_2=(x=7,p_x=15.4131)$ shown in the plane
$(z, p_z)$ (1st panel), $(z, p_y)$ (2nd panel), $(y, p_y)$ (3rd panel) and $(y, p_z)$ (4th panel)
for $E_0 = 9.8994$. Both cylinders were spanned by $173$
orbits, with initial conditions taken on a circle in the domain $(z, p_z)$ of the center
manifold. Here we fixed $\Lambda=0.001$, $E_r=10$, $\alpha_{21}=-100$, $A_2=-80$, $\alpha_{31}=10^{-4}$,
$\alpha_{32}=10^{-5}$, $A_3=10^{-5}$ so that $E_{cr} = 9.899494937274579$. $IS_1$ initial
conditions: $x=14.142135621521241$, $p_{x_0}=0$, $y_0=z_0$, $p_{y_0}=3p_{z_0}$.}
\label{ht3}
\end{figure*}
\par Another possibility is that the cylinders coalesce with each other: this rare
situation characterizes the absence of chaos in the model and in this sense the
dynamics is said to be regular/non-chaotic. Interestingly enough this is the case for the
parameter domain $(A)$ where the dynamics of the cylinders are regular as we show now.
We have not found any numerical evidence of the breaking of this regular behavior,
contrary for instance to the dynamics of Bianchi IX universes in bouncing braneworld
cosmologies\cite{maier,cqg}.
%
\par
To start let us consider the parameter configuration (\ref{CMM1}) used in part of the
numerical experiments of the previous sections. For these parameters
the saddle-center-center critical point $P_2$ is characterized by $E_{cr_2}=9.89949493727457$ and
$(x_{cr_2}=14.1421356215212,~p_x=0,~y=1,~p_y=0,~z=1,~p_z=0)$. The total energy of the system
is taken as $E_0=9.8994$ so that the energy available to the rotational degrees of freedom of
the center manifold is $(E_{cr_2}-E_0 \sim 10^{-4})$.
\par The results of the previous sections showed that two 4-dim cylinders, one stable $W_S$ and
one unstable $W_U$, both with the topology $R \times S^3$, emerge from a neighborhood of the center manifold about $P_2$.
The center manifold $S^3$ encloses the critical point $P_2$
and tends to it as $E_0 \rightarrow E_{cr_2}$. At this limit the cylinders $W_S$ and $W_U$ reduce to the
separatrix $S$ which makes a homoclinic connection to itself in the invariant plane. The separatrix is
a structure inside the cylinders, about which the flow with the oscillatory degrees of freedom $(y,p_y,z,p_z)$
proceeds guiding both cylinders towards the bounce. Their first encounter, with either a transversal crossing
or a smooth coalescence is expected to occur in a neighborhood of the bounce $(x_{b}=4.565245,p_x=0)$ where $x_b$
is the scale of the bounce for the orbits at $p_x=0$. In order to examine this first encounter
we will adopt as the surface of section\cite{licht} the 4-dim surface $\Sigma:(x=x_b,p_x=0)$.
\par
For the sake of numerical simplicity
here our simulations will be restricted to the dynamics on the two 4-dim
invariant submanifolds (\ref{eqInvSub1}) and (\ref{eqInvSub2}) of the 6-dim phase space which, in the
canonical variables (\ref{eqGen1}) and (\ref{eqGen2}), are expressed
\begin{eqnarray}
\label{eqInv3}
IS_1&:&~y=1,~p_y=0,\\
IS_2&:&~y=z,~p_y=3p_z.
\end{eqnarray}
\par In the first simulation we take $(x_0=x_{cr_2}, p_{x0}=0)$, and fix the initial conditions on the
4-dim invariant submanifold $IS_1$, namely, with $(y=1,p_{y}=0)$; such initial conditions are
obviously to be taken in the sector $(z,p_z)$ of the center manifold $S^3$,
which has the topology of $S^1$ and is defined by the Hamiltonian constraint (\ref{eq10091})
$H(x=x_{cr_2},p_x=0,y=1,p_y=0,z,pz,E_0=9.8994)=0$. By performing the evolution of $173$
initial conditions in the above set, the exact dynamics actually evolves a 4-dim invariant
subset $(x,p_x,z,p_z)$ of the full 6-dim phase space as expected due to our restriction to the
4-dim invariant submanifold $(y=1,p_y=0)$; in this particular simulation we have that, under
the exact dynamics, no motion is present in the sector $(y,p_y)$. We generate one 2-dim stable $W_S$
and one 2-dim unstable $W_U$ cylinders of orbits which initially move towards the first bounce.
As mentioned we adopt the surface of section $\Sigma_b:(x=x_b,p_x=0)$, where $x_b \simeq 4.565245$
is the scale factor of the bounce.
The points $(z_b,p_{z_b})$ resulting from the section of both cylinders ($W_S$ black
and $W_U$ gray) by the surface of section $\Sigma_b$ are displayed in Fig. \ref{pm1} (left),
corresponding actually the Poincar\'e maps of both cylinders on $\Sigma_b$.
This first Poincar\'e map at the bounce is a numerical evidence of the coalescence
of one cylinder into the other and gives a clear picture of the regular (non-chaotic)
motion in the dynamics of the cylinders.
We also display in Fig. \ref{pm1} (right) the Poincar\'e maps of the cylinders
in another surface of section $\Sigma_2=(x=7,p_x=15.413)$ showing also the
coalescence of the two cylinders into one another in a time after the first bounce.
\par From these Poincar\'e maps we see the coalescence of the stable cylinder (black dots) and unstable
cylinder (gray diamonds) in two sections in the phase space.
In fact, it can be shown that this coalescence is maintained for any section
of the phase space crossed by orbits in the stable or unstable manifold.
This is a integrability signature of the dynamics showing a feature
of no chaos in the model. Although the numerical simulations shown here
were done for the parameters (\ref{CMM1}) we have checked that this integrability
pattern is maintained in general for all parameter configurations of the domain $(A)$ in which $P_1$
is a center-center-center and $P_2$ is a saddle-center-center provided by a proper choice
of the coupling constants in the potential $U_{HL}$.
\par
To complete our analysis we have also considered the case of the second invariant submanifold $IS_2$.
Again we obtain here numerical evidence of the regularity of the dynamics as given in Figs. \ref{ht2} and \ref{ht3} .
Here we plot the Poincar\'e maps in the surface of section $\Sigma_b:(x=x_b,p_x=0)$
at the first bounce (corresponding to the points $(y_b, p_{y_b}, z_b, p_{z_b})$)
of the stable cylinder (black dots) and the unstable cylinder (gray diamonds)
as shown in Figure 5, and in the surface of section $\Sigma_2=(x=7,p_x=15.4131)$.
corresponding to the points $(y_b, p_{y_b}, z_b, p_{z_b})$ as shown in Fig. 6.
These Poincar\'e maps show clearly the coalescence of the stable cylinder (black dots)
and unstable cylinder (gray diamonds) in two arbitrary sections in the phase space,
in common with the case of the first invariant submanifold $IS_2$.
As in the previous experiments we verified numerically that this coalescence is maintained
for any surface of section of the phase space crossed transversally by the stable and the unstable manifolds.
This is a regular signature of the dynamics showing a feature
of no chaos in the model. Furthermore we also checked that this regular (non-chaotic)
pattern of the dynamics, obtained for the invariant submanifold $IS_2$ with the
parameters (\ref{CMM1}), is maintained in general for parameter configurations $(A)$ in which $P_1$
is a center-center-center and $P_2$ is a saddle-center-center.
\par The patterns of the phase space dynamics of a general Bianchi IX cosmological model
discussed in the previous sections are fundamentally connected to the
general potential $U_{HL}$ (\ref{eq5}) of a non-projectable version of Horava-Lifshitz gravity
which, among other characteristics, allows for the presence of nonsingular bounces in the orbits
of the model due to curvature dependent potentials.

The rich dynamics of the model is mainly due to the number of parameters introduced
via the HL potential which in turn demands a careful
classification of the pairs of critical points in the finite region of the phase space.
In the domain of parameters $(A)$ examined in sections \ref{sectiona}, \ref{centermanifold} and  \ref{chaosINV}
the critical points are a center-center-center and a saddle-center-center so that the phenomena
in phase space are of the same nature of the ones discussed in \cite{ozorio,cqg,maeda}.
Now in the following two sections we will examine the parameter domains $(B)$ and $(C)$
in which some features of the phase space dynamics -- mainly connected to the
presence of a saddle with multiplicity two -- are new and, to our knowledge,
not yet seen in the literature.


\section{A saddle of multiplicity two and the p{$_Z$}-momentum attractors}

\begin{figure}
\includegraphics[width=6cm,height=6cm]{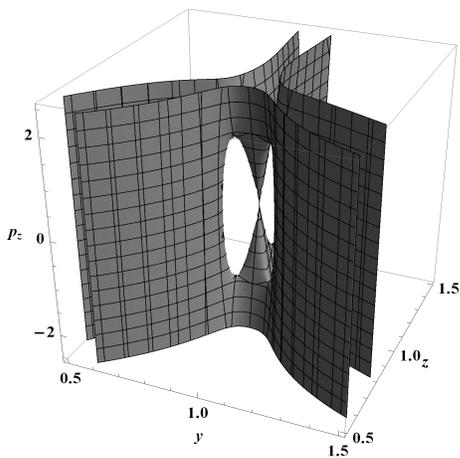}
\caption{Numerical illustration of the phase space in a neighborhood of
the center-saddle-saddle $P_1$, corresponding to the parameter configuration $(\ref{CMM2})$.
Here we display the section $(x=x_{cr_1}=7.071067829543,~p_x=0,~p_y=0)$ of the 5-dim
phase space for $E_0=E_{cr_1}$. The critical point $P_1$
is located at the common vertex of the cones into which the $4$-hyperboloid degenerates
for $E_0 = E_{cr}$. The high instability of the dynamics outside the invariant plane is due to
the presence of a saddle of multiplicity two at $P_1$.}
\label{htLL}
\end{figure}

Here our focus will be in the parameter domain $(B)$
where $P_1$ is a center-saddle-saddle and $P_2$ is a saddle-center-center.
The topology of the phase space in the neighborhood of $P_1$ has the structure of a
saddle of multiplicity two times $S^1$, since $P_1$ in $(B)$ has $q<0$ and $q_x < 0$ (cf. (\ref{eqHC})).
This topological feature induces a high instability in the phase space dynamics.
In order to better grasp such behavior we illustrate in Fig. \ref{htLL}
the topology of the phase space about $P_1$ with the parameters
\begin{eqnarray}
\label{CMM2}
\nonumber
\Lambda&=&0.001,~~~E_r=1,\\
\alpha_{31}&=&10^{-4},~~~\alpha_{32}=10^{-5},~~~\alpha_{33}=-0.00092,\\
\nonumber
\alpha_{21}&=&0,~~~\alpha_{22}=-248/3,
\end{eqnarray}
with $x_{cr_1}=7.071067829543$ and $E_{cr_1}=9.192388160728$.
While on the invariant plane the motion about $P_1$ is that of periodic orbits,
outside the invariant plane the hyperbolic motion is the origin
of a high instability in the phase space dynamics.
Fig. \ref{htLL} displays the phase space section ($x=x_{cr1}$, $p_x=0$, $p_y=0$)
of the hyperbolic motion about $P_1$. In the figure we note
that the critical point $P_1$ is located at the common vertex of cones
into which the $4$-hyperboloid degenerates for $E_0=E_{cr1}$.
\par Let us now discuss some new features in the dynamics of orbits visiting a
nonlinear neighborhood of $P_1$ connected with its character of a saddle of multiplicity
two.
\par For the parameters (\ref{CMM2}) adopted, let us consider the saddle-center-center critical
point $P_2:(x=x_{cr_2}=14.14213562152124,p_x=0,y=1,p_y=0,z=1,p_z=0)$
with critical energy $E_{cr_2}=9.89949493727457$. We construct
the unstable cylinder $W_U$ (gray) that emerges from the neighborhood of  $P_2$ towards the bounce,
spanned by $26$ orbits, with initial conditions taken on a circle in the domain $(z,p_z)$
of the center manifold about $P_2$ with energy $E_0=9.8994$.
From the same initial conditions we generate the stable cylinder $W_S$ (black) that
also emerges towards the bounce.
\begin{figure}
\includegraphics[width=6cm,height=6cm]{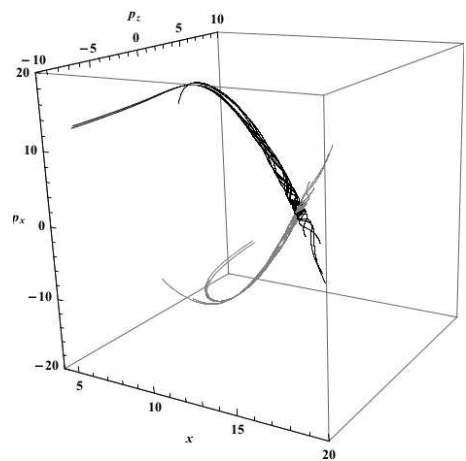}
\caption{Numerical illustration of the unstable cylinder (gray) and stable cylinder (black),
that emerge from a neighborhood of $P_2$ towards the bounce,
spanned each by $26$ orbits with initial conditions taken on a circle in the domain $(z, p_z)$
of the center manifold about $P_2$ for $E_0 = 9.8994$, corresponding to the parameters (\ref{CMM2}). The
projection of the figure in the plane $(x, p_x)$ ``shadows'' the separatrix of the invariant
plane. Due to the instability in the dynamics connected to the
saddle with multiplicity two in $P_1$,
part of the orbits escape to two additional $p_z$-momentum attractors
with an infinitely large anisotropy parameter, at
$x={\rm const}$, $z={\rm const}$ and $p_z \rightarrow \pm \infty$, as shown
in the figure.
}
\label{htLL1}
\end{figure}
These cylinders are illustrated in Fig. \ref{htLL1} from where three distinct sets of orbits
can be singled out. According to the dynamics examined in previous sections, these orbits would be expected
to have two attractors, either the center manifold itself or the de Sitter attractors at infinity.
These two set of orbits (connected with the deSitter or the center manifold attractors) are
seen in the Figure, the projection of which on the invariant plane $(x, p_x)$ ``shadows'' the separatrix
of the invariant plane.
\par
However in the present case $(B)$, due to the high instability connected to the
saddle-saddle-saddle $P_1$, we observe a third set of orbits that visit a
nonlinear neighborhood of $P_1$ and escape to two additional $p_z$-momentum attractors
with a very large anisotropy parameter at $x={\rm const.}$, $z={\rm const.}$
and $p_z \rightarrow \pm \infty$ as can also be seen in Fig. \ref{htLL1}.
Due to the high instability of the dynamics of these orbits, the numerical
evaluation for long times is quite critical, demanding an accuracy which is
in the available limit of the codes used in this work. Let us consider, for instance, an orbit
on the unstable cylinder belonging to this third set, generated from the initial conditions
\begin{eqnarray}
\label{ICshear}
\nonumber
x&=&x_{cr2},~~p_x=0,~~y=1,~~p_y=0,\\
z&=&1.00075,~~p_z=0.5954540145554457,
\end{eqnarray}
taken on the circle $(z,p_z)$ of the section $(y=1,p_y=0)$ of the center manifold about $P_2$,
for $E_0=9.8994$. We kept the dynamics restricted to the time interval $t=[0,477.39]$ so that
the Hamiltonian constraint ${\cal H}$ (\ref{eq10091}) is still conserved, namely, ${\cal H} \leq 3.337 \times 10^{-12}$.
At $t=477.39$ we obtain $x \simeq 4.51755$, $z \simeq 0.201023$
and $p_z \simeq 910.195$, leading to a value of the anisotropy parameter $\sigma^2 \simeq 3.93864$
which is larger than the initial anisotropy by eight orders of magnitude.
For $t$ slightly larger than $t_f=477.39$ the conservation of ${\cal H}$ breaks up, with the
value of $p_z$ increasing exponentially. We therefore conclude that the asymptotic
configuration of the orbits of the third set, shown in Fig. \ref{htLL1}, escape to
two $p_z$-momentum attractors with an infinitely large anisotropy parameter, at
$x={\rm const}$, $z={\rm const}$ and $p_z \rightarrow \pm \infty$.
This is a direct consequence of the dynamical instability associated with the saddle
of multiplicity two at $P_1$, a feature not yet observed in the results of the previous Sections.
\par Finally we must note that in our present example the
dynamics of the orbits is restricted to the invariant submanifold $(y=1,p_y=0)$, so that
the anisotropy parameter reduces to
$\sigma^2= \Big({z^2 p_z^2}/x^6 \Big)$, cf. (\ref{shear1}).

\section{The center manifold about a saddle of multiplicity two: parametric bifurcation\label{bifurc}}

Finally we discuss here the phase space dynamics corresponding to
a system whose parameter configuration is in the domain $(C)$. As we will see the skeleton of the dynamics is
dominated by the saddle-saddle-saddle critical point $P_2$, and the effect of the saddle with multiplicity two
on the dynamics about $P_2$ is examined. We should mention that the presence of a saddle of multiplicity two in
physical systems is rare (possibly absent in the case of cosmological models) and therefore we
are led to undertake a more detailed examination of this case.
\begin{figure*}
\includegraphics[width=8cm,height=6cm]{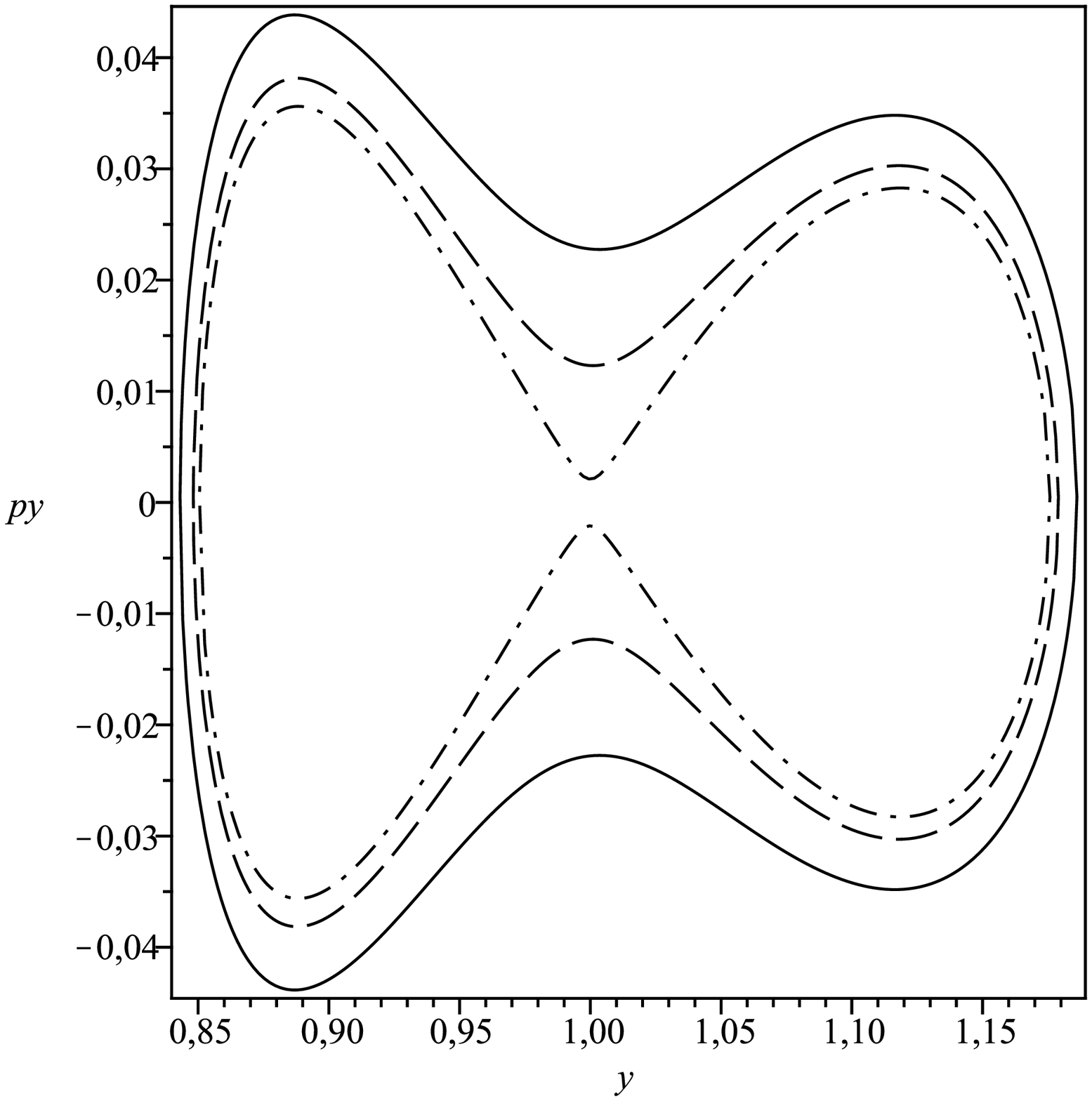}~~\includegraphics[width=8cm,height=6cm]{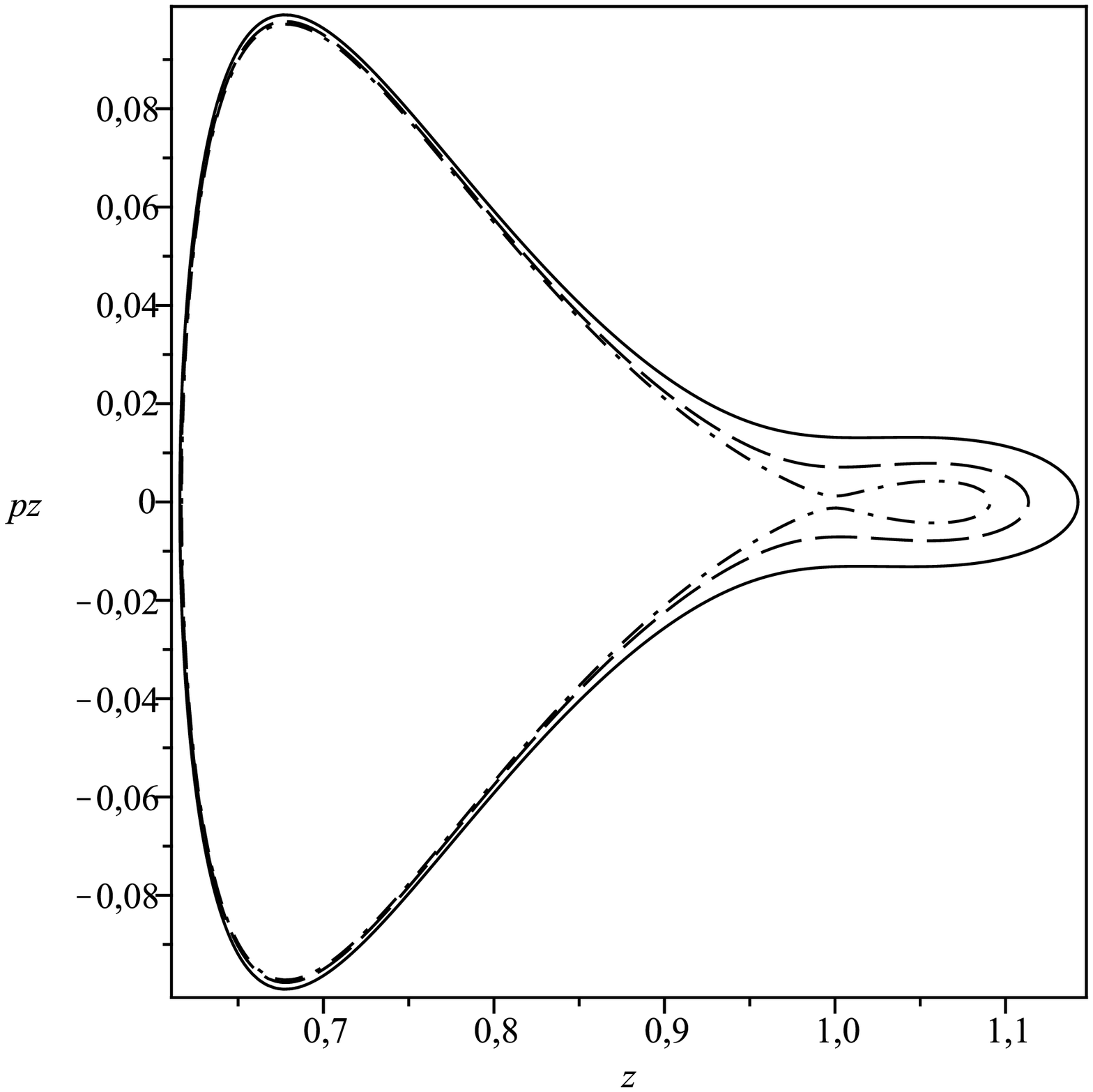}
\caption{Section $(p_z=0,z=1)$ (left) and section $(p_y=0,y=1)$ (right) of the 3-dim center manifold (\ref{CMH3})
for $E_0=0$ (solid),~$0.00001$ (dashed) and $0.000014$ (dash-dotted), corresponding to
the parameter configuration (\ref{eqSS2}) with $q<0$. For these energies $E_0<E_{cr_2}$ the manifold is topologically a 3-sphere
enclosing the critical point $P_2$ which is a saddle-saddle-saddle. As $E_0 \rightarrow E_{cr_2}$ the curves
pinch at the critical point $P_2$, corresponding to $(y=1,p_y=0)$ (left figure) and $(z=1,p_z=0)$ (right figure).
}
\label{Section1}
\end{figure*}
Let us consider the following parametric configuration in $(C)$
\begin{eqnarray}
\label{eqSS2}
\nonumber
\Lambda&=&5/60,~~~Er=0,\\
\alpha_{31}&=&0.1,~~~\alpha_{32}=0,~~~\alpha_{33}=1,\\
\nonumber
\alpha_{21}&=&1.4601,~~~\alpha_{22}=1/3,
\end{eqnarray}
for which we obtain the two critical points $P_1$ and $P_2$ with
\begin{eqnarray}
\label{eqSS3}
\nonumber
x_{cr_1}&=&0.715744514081549,\\
\nonumber
 x_{cr_2}&= &2.09499845891933,
\end{eqnarray}
corresponding to the two positive real roots of equation (\ref{eq11}).
For $P_1$ we obtain that the energy $E_{cr_{1}}<0$ so
that this critical point is out of the physical phase space.
Therefore in the parameter configuration (\ref{eqSS2}) the physical system has
only one critical point $P_2$, the energy of which is
\begin{eqnarray}
\label{eqSS4}
E_{cr_{2}}= 0.00001411994285.
\end{eqnarray}
For $P_{2}$ we also evaluate that
\begin{eqnarray}
\label{eqSS5}
q= -0.00267275969793, ~q_x = 1.32604264858754,
\end{eqnarray}
characterizing $P_{2}$ as a saddle-saddle-saddle
(where a saddle with multiplicity two is present, cf. (\ref{eqHH})).
The critical point $P_2$, which is denoted
a saddle-saddle-saddle, corresponds actually to the topological product of a saddle
times a saddle with multiplicity two. Therefore, to avoid a saturation in the remaining text,
we will sometimes refer to $P_2$ simply as a saddle with multiplicity two.
We now proceed to examine the topology of the phase space about this critical point.
\par
To start let us examine the possible linear motions about $P_{2}$.
Let us consider the case $E_x=0$, cf. (\ref{eqHC3}). The first possibility
corresponds to $(x=x_{cr_2},p_x=0)$ implying that the motions are orbits on the
3-dim surfaces
\begin{eqnarray}
\label{eqCM1}
\nonumber
\Big[\frac{1}{2}\frac{p_y^2}{ M_0^3} - 3 |q| (y-1)^2\Big]
+\Big[\frac{3}{2} \frac{p_z^2}{M_0^3}- |q| (z-1)^2\Big]\\
=2 (E_{cr_2}-E_0),
\end{eqnarray}
which depend continuously on $E_0$. For $(E_{cr_2}-E_0)$
sufficiently small so that (\ref{eqCM1}) holds, these constant energy surfaces
have the structure of a $3$-hyperboloid with the constant of motions $C_1$, $C_2$ and $C_3$ satisfying the algebra of the
three-dimensional hyperboloid group under the Poisson bracket operation (cf. (\ref{eqHC6})
with $q<0$). However, contrary to the cases of the previous sections where the $S^3$ center manifold reduces to
the critical point as $E_0 \rightarrow E_{cr_2}$, here the $3$-hyperboloid invariant manifold consists of
the critical point from which emanate the saddle lines $p_y=\pm \sqrt{3} \mu ~ (y-1)$ and
$p_z=\pm (\mu/\sqrt{3})~(z-1)$ with multiplicity two, where $\mu=\sqrt{2|q|M_{0_2}^3}$.
Actually in the case of $E_0=E_{cr_2}$ the hyperbolic phase space dynamics about a
neighborhood of the critical point is analogous to that illustrated in Fig. \ref{htLL},
where there $3$-hyperboloid degenerates into two $3$-cones with a common vertex at the critical point.
The nonlinear extension of the center manifold, obtained as $(E_{cr_2}-E_0)$ increases,
exhibits a rich structure connected to the presence of a saddle with
multiplicity two and its bifurcations, as we now proceed to discuss.
\par For the configurations analyzed
in the previous sections, for which $P_2$ had $q>0$, the existence of the center manifold demanded that
$E_{cr_2}-E_0 \geq 0 $; for the equality case the center manifold reduced to a point,
the saddle-center-center critical point.
Now since $q<0$ this restriction no longer exists as can be clearly seen
from (\ref{eqCM1}). Our analysis will contemplate separately
the following energy domains,
\begin{eqnarray}
\label{eqNO}
\nonumber
(I):~~&& (E_{cr_2}-E_0) > 0,\\
(II):~~&&  E_{cr_2}-E_0 = 0,\\
\nonumber
(III):~~&& E_{cr_2}-E_0 < 0.
\end{eqnarray}
\begin{figure*}
\includegraphics[width=6cm,height=5cm]{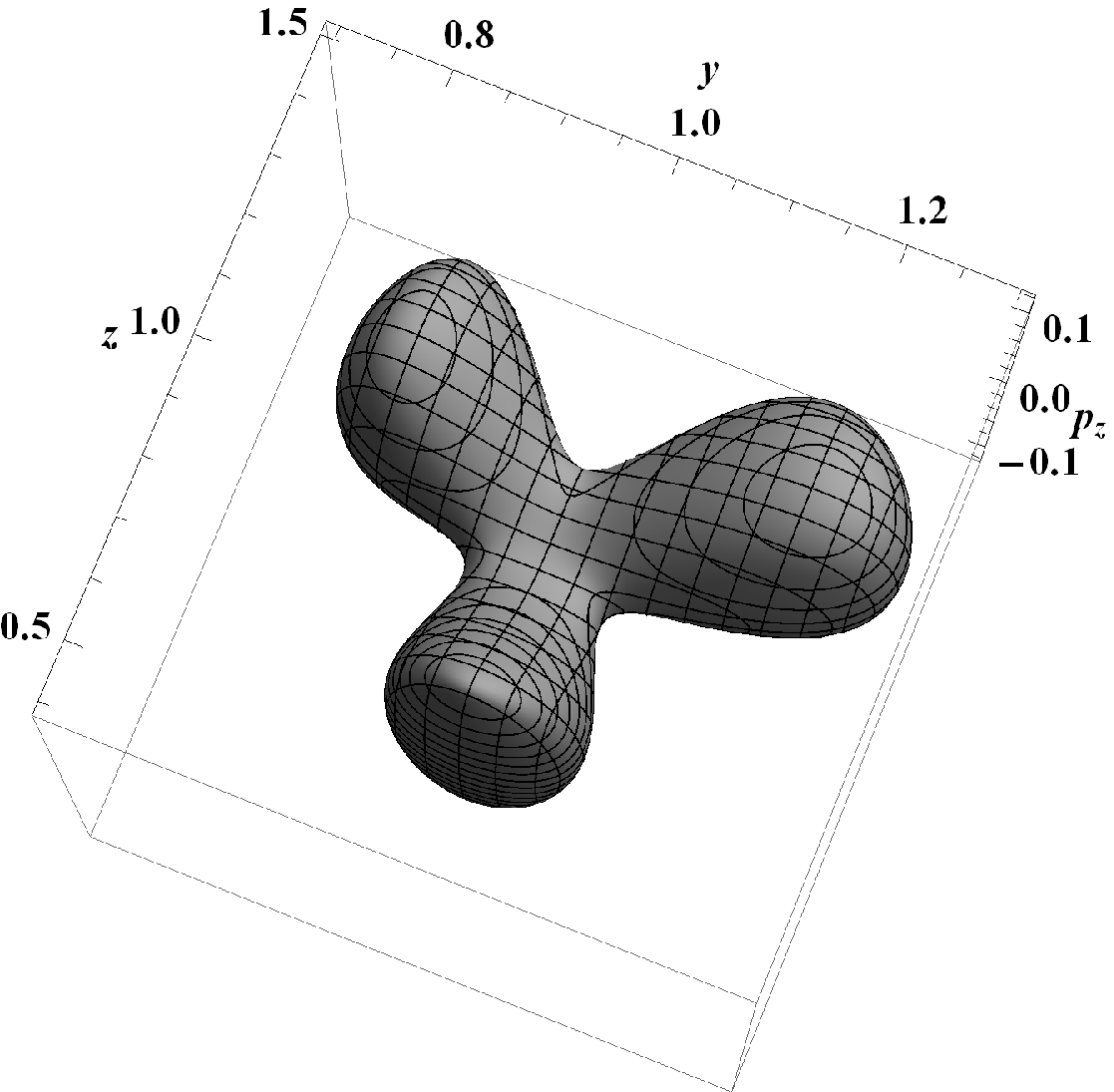}\includegraphics[width=6cm,height=5cm]{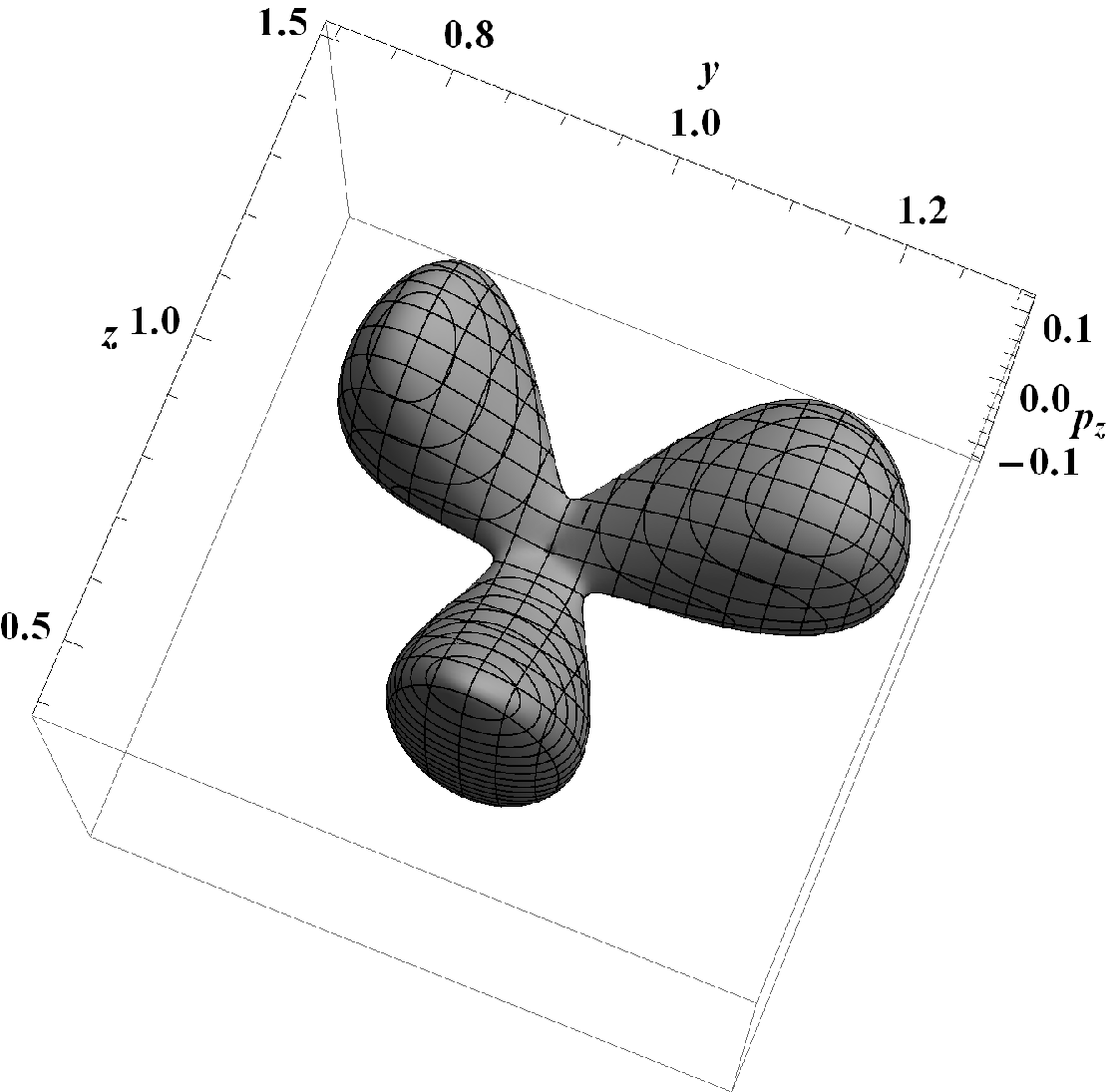}\\
\includegraphics[width=6cm,height=5cm]{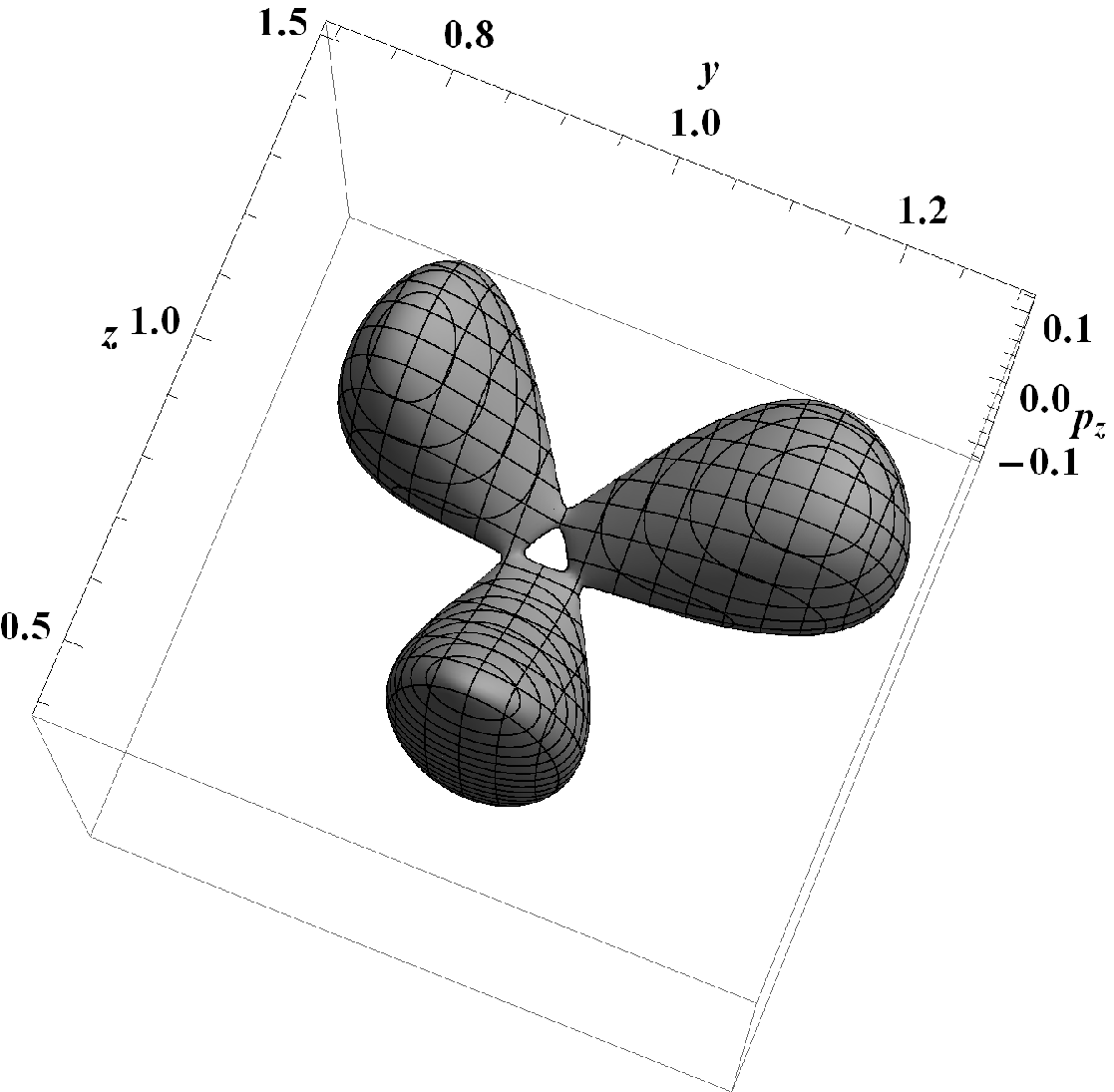} \includegraphics[width=6cm,height=5cm]{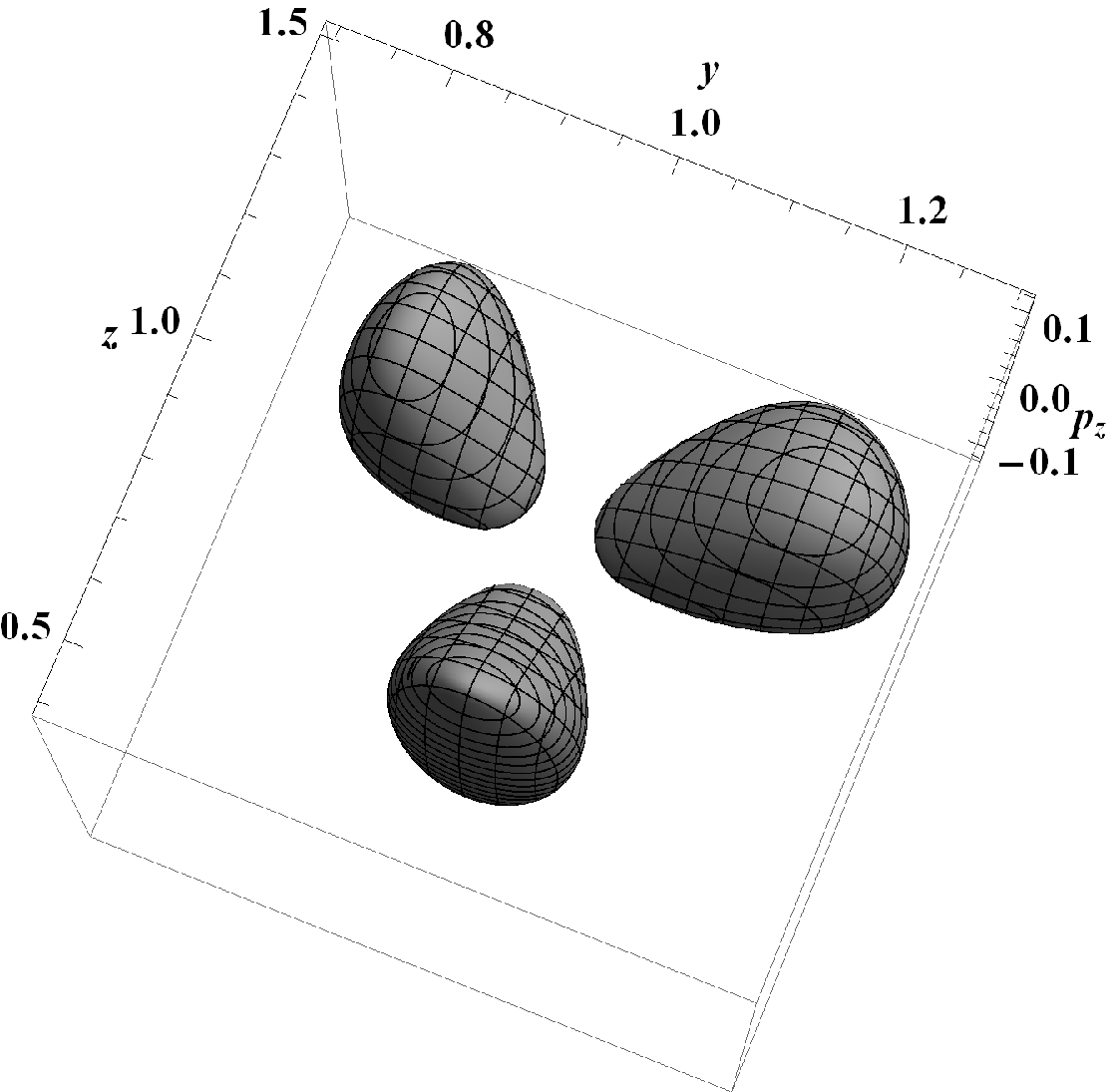}
\caption{The 2-dim sections $p_y=0$ of the 3-dim center manifold (\ref{CMH3}) for several significative energies:
(i) $E_0=0$ (top,~left): (\ref{CMH3}) has the topology of a $S^3$ enclosing the saddle with multiplicity two, a
pattern which holds for the domain $0 \leq E_0 < E_{cr_2}$. (ii) $E_0=E_{cr_2}$ (top,~right): the saddle with multiplicity two now belongs
to (\ref{CMH3}), being  the one point connection of two leaves of the manifold and breaking the $S^3$ topology, so that this manifold
contains infinitely many homoclinic orbits. (iii) $E_0=0.0000153> E_{cr_2}$ (bottom,~left): showing the bifurcation of $S^3$
into a 3-dim torus. The critical point is outside the $3$-torus. (iv) for larger values, $E_0=0.000025$ (bottom,~right):
the center manifold becomes multiply connected. The parameters of this configuration are given in ($\ref{eqSS2}$).
}
\label{birf1}
\end{figure*}
\par In the energy domains ({\ref{eqNO}}) the $3$-dim center manifold about $P_2$ (containing a saddle
of multiplicity two) is given by the Hamiltonian constraint,
\begin{eqnarray}
\label{CMH3}
\nonumber
{\cal H}_{C}&=& \frac{p_y^2 y^2}{2 x_{cr_2}^3} + \frac{3 p_z^2 z^2}{2 x_{cr_2}^3}+\frac{x_{cr_2}}{2 z^{\frac{4}{3}}}
- \frac{x_{cr_2}}{y z^\frac{1}{3}} - \frac{x_{cr_2} y}{z^\frac{1}{3}} \\
\nonumber
&-& x_{cr_2} z^\frac{2}{3} +
\nonumber
\frac{x_{cr_2} z^\frac{2}{3}}{2 y^2} + \frac{1}{2} x_{cr_2} y^2 z^\frac{2}{3} + 2 x_{cr_2}^3 \Lambda \\
&+& 2 E_0 + \frac{2 E_r}{x_{cr_2}}+{x_{cr_2}^3}U_{HL}(x_{cr_2},y,z)=0.
\end{eqnarray}
\par
In the energy domain $(I)$  the $3$-dim surface (\ref{CMH3}) is a topological $3$-sphere enclosing
the saddle with multiplicity two $P_2$, as illustrated in Figs. \ref{Section1}, where we plot
its sections $(p_z=0,z=1)$ (left) and section $(p_y=0,y=1)$ (right),
for $E_0=0$ (solid),~$0.00001$ (dashed) and $0.000014$ (dash-dotted). As $E_0=E_{cr_2}$ the curves
pinch at the critical point $P_2$, corresponding to $(y=1,p_y=0)$ (left figure) and $(z=1,p_z=0)$ (right figure).
A larger dimensional view of this case is also illustrated in Fig. \ref{birf1}, (top,~left),
where we plot the 2-dim section $p_y=0$ of the invariant 3-dim center manifold (\ref{CMH3}) for $E_0=0$, showing
a topological $S^3$ (with three lobes) enclosing the saddle with multiplicity two.
\par
As $E_0$ increases towards the value $E_{cr_2}$ the closed surface deforms,
reaching the domain $(II):~(E_0=E_{cr_2})$ when the closed surface {\it pinches} at the critical point
$M_0=(x=x_{cr_2},p_x=0,y=1,p_y=0,z=1,p_y=0)$, as shown in Fig. \ref{birf1}, (top,~right). In this case
the center manifold is said to undergo a bifurcation;
the saddle with multiplicity two $P_2$ now belongs to (\ref{CMH3}), being  a common point of the lobes
and breaking the $S^3$ topology, so that this manifold
contains infinitely many homoclinic orbits to the critical point.
\par As the energy parameter enters the domain $III:~E_0>E_{cr_2}$ a further bifurcation occurs
and the $3$-dim manifold becomes topologically a $3$-torus, as illustrated in Fig. \ref{birf1} (bottom,~left),
showing the bifurcation of a $S^3$ into a $3$-torus. The critical point is outside the $3$-torus.
Finally, for larger values of $E_0$ in the domain $(III)$
the invariant center manifold becomes multiply connected, Fig. \ref{birf1}, (bottom,~right).
\par The above picture is in bold contrast with the case of a center-center with multiplicity two,
examined in Sections \ref{sectiona}-\ref{centermanifold}, where the 3-dim center manifold about
this critical point is defined for $0 \leq E_0 \leq E_{cr_2}$ and
reduces to a single point (the critical point) for $E_0=E_{cr_2}$.
For $E_0>E_{cr_2}$ the center manifold does not exist.
\par A more extended examination of these structures and the associated features of the whole
phase space dynamics, as well as its implications to Cosmology, is beyond the scope of the present
paper and will be dealt with in a future work. However we should mention that, to our knowledge,
such features have not yet been seen in the literature of cosmological models.

\section{Final Comments and Conclusions}

In this paper we examined the phase space dynamics of general bouncing Bianchi IX cosmological
models in which nonsingular bounces are generated by extra higher order spatial curvature terms
in the framework of the Ho\v{r}ava-Lifshitz (HL) gravity. The HL gravity action adopted contains five
independent parameters, apart from the $\lambda$ parameter that breaks the invariance under
four-dimensional diffeomorphisms present in classical General
Relativity. In order to recover the classical regime, our analysis was restricted to $\lambda=1$.
In the $U_{HL}(^{(3)}R)$ potential considered in the paper the five independent parameters
were restricted by imposing $A_3>0$ (cf. (\ref{eqC2})) so that the dynamics is nonsingular
implementing instead bounces in the dynamics of the model.
Furthermore in the class of models analyzed we have restricted ourselves to an energy-momentum
tensor of dust and radiation, which are conserved independently, plus a positive cosmological constant $\Lambda$.
The corresponding total energy of dust turns out to be  a constant of motion connected to the
total conserved Hamiltonian.
As a consequence of $A_3>0$ and $\Lambda>0$ the model contains just two critical points
$P_1$ and $P_2$ at the the finite region of the phase space. The nature of these critical points
determines the structure of the phase space dynamics and of the attractors at infinity.
\par
Our treatment in the paper is based strongly on the Hamiltonian formulation,
with a conserved Hamiltonian constraint plus the associated Hamilton's equations of motion.
By the use of appropriate canonical variables we were able to make a global
examination of the structures of the phase space that organize the dynamics,
as critical points, center manifolds, homoclinic cylinders emanating
from the center manifolds and the attractors at phase space infinity.
\par
\par In Section \ref{section3} the nature of the critical points in the $6$-dim phase space was examined
by the linearization of Hamilton equations about these points.
In this context we were able to classify the dynamics in four distinct parameter domains according
to the possible nature of the critical points: domain $(A)$, where the critical point $P_1$ is a
center-center-center and the critical point $P_2$ is a saddle-center-center; domain $(B)$, where $P_1$
is a center-saddle-saddle and $P_2$ is a saddle-center-center; domain $(C)$, where $P_1$
is a center-saddle-saddle and $P_2$ is a saddle-saddle-saddle; domain $(D)$, where
$P_1$ is a center-center-center and $P_2$ is a saddle-saddle-saddle.
In all four domains, with its respective structures of critical
points and of attractors at infinity, the Bianchi IX models are non-singular
in the sense that the spacetime curvature does not diverge and the physical average scale factor $x(t)$
never reaches zero. All phase space orbits discussed in the paper are either periodic
(perpetually bouncing solutions), or oscillatory with an eventual escape to one of the
attractors at the infinity of phase space, or orbits homoclinic to a center manifold.
\par The features of the parameter domain $(A)$ were examined in Sections \ref{sectiona}-\ref{chaosINV}.
The critical points are a center-center-center $P_1$ and a saddle-center-center $P_2$.
We introduced a new set of canonical variables $(x,p_x,y,p_y,z,p_z)$ that separate
the degrees of freedom of the system into two rotational modes $(y,p_y)$ and $(z,p_z)$,
about a linear neighborhood of the center-center sector of both critical points and
an expansion/contraction mode $(x,p_x)$ along the saddle direction of $P_2$ or
a further rotational mode along the additional center direction of $P_1$.
The rotational modes for both critical points, connected to the presence of a center of multiplicity two,
are defined on the {\it center manifold} of unstable periodic orbits which has the
topology $S^3$. A necessary condition for the existence of the center manifold is $(E_{cr}-E_0) >0)$,
where $E_{cr}$ is the energy of the critical point and $E_0$ is the total energy of the system.
By continuity as $(E_0-E_{cr})$ increases the nonlinear extension of the center manifold maintains the topology
of $S^3$. In the case of $P_2$, together with the saddle variables $(x,p_x)$ it defines the 4-dim stable
and unstable cylinders, with topology $R \times S^3$ that coalesce to the center manifold as
$t \rightarrow \pm \infty$ respectively.
Summing up, the topology of the phase space about the center-center-center $P_1$ is $S^1 \times S^3$ and about
the saddle-center-center $P_2$ is $R \times S^3$. Therefore in a neighborhood of $P_2$ the variables $(x,p_x)$
have a saddle nature, while in the neighborhood of $P_1$ they have a rotational nature.
With the use of the canonical variables $(y,p_y,z,p_z)$ we obtain an exact analytical form for
the center manifold as well as an accurate numerical description of the phase space dynamics
in extended regions away from the critical points.
\par In Section \ref{centermanifold} we then considered two characteristic types of orbits obtained
from distinct sets of parameters in the domain $(A)$ and appropriate initial conditions
on the center manifold. The first case corresponds to perpetually bouncing periodic orbits orbits
(propagated forward or backward in time) and the second case corresponds to oscillatory orbits
that undergo a finite number of bounces before escaping to the deSitter attractor at infinity.
In both cases the frequency of the oscillatory modes of the orbit increases substantially
as it visits the neighborhood of the bounces.
We also examined the evolution of the anisotropy parameter which is connected with
the rotational mode variables. We obtained that
the anisotropy is oscillatory and bounded, increasing several orders of magnitude as the orbits
visit a neighborhood of the bounce. In particular the anisotropy of the second set of orbits, with
a finite number of bounces, goes to zero as the orbits reach the deSitter attractor. This parameter
will be useful in the recognition of the nature of anisotropic momentum attractors that appear in
the domain parameter $(B)$.
Also, for the parameter domain $(A)$, we examined in Section \ref{chaosINV} the question of regular and/or chaotic motion connected to
the $4$-dim homoclinic cylinder structures emanating from the center manifold about
$P_2$. Contrary to results of homoclinic chaos originated from the transversal crossings of homoclinic cylinders
in Bianchi IX bouncing brane cosmologies\cite{maier,cqg}, in the present set of
Ho{\v r}awa-Lifshitz Bianchi IX bouncing cosmologies we obtained the smooth coalescing
of the stable and the unstable homoclinic cylinders, characterizing thus a regular
dynamics in the invariant submanifolds of the model. This is a rare result of the
regularity of the dynamics in the presence of homoclinic cylinders, not yet seen in the
literature of cosmological models.
\par Completely new distinct dynamical patterns appear in connection with the
critical points in the parameter domains $(B)$ and $(C)$. In the case of $(B)$ the
two critical points are a center-saddle-saddle $P_1$ and a saddle-center-center $P_2$.
As discussed in detail in Sections \ref{sectiona}-\ref{centermanifold}, in a
neighborhood of a saddle-center-center $P_2$ we have the general pattern of stable and
unstable cylinders of orbits emanating from the center manifold about $P_2$.
However in the present case the critical point $P_1$ contains a saddle of multiplicity two
which is the source of a high instability in the dynamics, acting on the cylinders
as they visit a nonlinear neighborhood of $P_1$.
In fact our numerical experiments showed that orbits of stable and unstable cylinders emerging
from the center manifold about $P_2$ towards the bounce can be classified in three distinct sets
according to their attractors: (i) orbits that have the center manifold as an attractor, (ii)
orbits that have the de Sitter configurations at infinity as an attractor and (iii) the two further
attractors at $p_z \rightarrow \pm \infty$. The momentum attractors (iii) correspond to a configuration
of infinite anisotropy; this is a direct consequence of the dynamical instability associated with the saddle
of multiplicity two at $P_1$, a feature not observed in the results of the previous Sections.
\par In Section \ref{bifurc} we discussed the properties of the center manifold about
the saddle-saddle-saddle critical point $P_2$, in the case of the parameter configuration $(C)$.
The presence of a saddle of multiplicity two in $P_2$ engenders a rich structure in the phase space
not yet observed in the literature. Contrary the previous cases -- where the $3$-dim center manifold
is defined for $0 \leq E_0 \leq E_{cr_2}$ only and reduces to a single point (the critical point)
for $E_0=E_{cr_2}$ -- the center manifold is defined for all $E_0 \geq 0$ and undergoes bifurcations
with increasing $E_0$. For $E_0 < E_{cr_2}$ the topology of the center manifold is $S^3$ enclosing the
critical point $P_2$. For $E_0=E_{cr_2}$ it turns into a $S^3$ with two points identified with $P_2$.
In this case the center manifold contains infinitely many orbits homoclinic to the critical point $P_2$.
For $E_0 >E_{cr_2}$ the manifold turns into a topological torus. Finally for $E_0$ sufficiently large
the center manifold becomes multiply connected with three distinct lobes.
The dynamics in the whole phase is highly unstable and its detailed examination is
beyond the scope of the present paper. It will eventually be discussed in a future publication.
\par Finally the fourth parameter domain $(D)$ was not examined since most of its features are
present already in the other domains.

\section*{Acknowledgements}

The authors acknowledge the partial financial support of CNPq/MCTI-Brazil. The Figures were generated
using the Wolfram Mathematica $7$ and the Dynamics Solver packet \cite{ds}.

\end{document}